\documentclass[preprint,12pt]{elsarticle} % preprint/review

\usepackage{lineno}
\modulolinenumbers[5]
\usepackage{color}
\usepackage[usenames,dvipsnames]{xcolor}
\usepackage{fullpage}
\usepackage{amsmath}
\usepackage{amssymb}
\usepackage{mathrsfs}
\usepackage{newfloat}
\usepackage{graphicx}
\usepackage{stmaryrd}
\usepackage{subfig}
\usepackage{multirow}
\usepackage{stackengine}
\usepackage{caption}
\usepackage{tikz}
\usepackage{multicol}
\usetikzlibrary{arrows,positioning,calc}
\usepackage{pstool}
\usepackage{textcomp}
\usepackage[hidelinks]{hyperref}
\hypersetup{colorlinks,bookmarksopen,linkcolor=blue,citecolor=blue,urlcolor=blue}
\usepackage{float}
\usepackage{makecell}
\usepackage[export]{adjustbox}
\usepackage{pdfpages}

\definecolor{myred}{RGB}{222,45,38}
\definecolor{myblue}{RGB}{0,115,189}
\definecolor{mygreen}{RGB}{49,156,54}

\newcommand{\bs}[1]{{\boldsymbol{#1}}}

\DeclareFloatingEnvironment{algorithm}
\DeclareRobustCommand\sampleline[1]{%
	\tikz\draw[#1] (0,0) (0,\the\dimexpr\fontdimen22\textfont2\relax)
	-- (2em,\the\dimexpr\fontdimen22\textfont2\relax);%
}

\newcommand{\tensorfour}[1]{{{^4}\boldsymbol{#1}}}
\newcommand{\column}[1]{{\underline{#1}}}
\pgfdeclarelayer{bg} % declare background layer
\pgfsetlayers{bg,main} % set the order of the layers (main is the standard layer)
\graphicspath{{figures/}}

\journal{J. Appl. Mech.}

\biboptions{authoryear}
\begin{document}
	
%\pagecolor{lightgray}

\begin{frontmatter}
\title{Level set based eXtended finite element modelling of the response of fibrous networks under hygroscopic swelling\tnoteref{titlefoot}}
\tnotetext[titlefoot]{The post-print version of this article is published in \emph{J. Appl. Mech.}, \href{https://doi.org/10.1115/1.4047573}{10.1115/1.4047573}}

\author[TUeM,ULB,M2i]{P.~Samantray} % \fnref{myfootnote}
\ead{priyam.samantray@gmail.com}

\author[TUeM]{R.H.J.~Peerlings} % \fnref{myfootnote}
\ead{R.H.J.Peerlings@tue.nl}

\author[TUeB]{E.~Bosco}
\ead{E.Bosco@tue.nl}

\author[TUeM]{M.G.D.~Geers}
\ead{M.G.D.Geers@tue.nl}

\author[ULB]{T.J. Massart}
\ead{thmassar@batir.ulb.ac.be}

\author[TUeM]{O.~Roko\v{s}\corref{correspondingauthor}}
\ead{O.Rokos@tue.nl}

% TU/e mechanical engineering address
\address[TUeM]{Mechanics of Materials, Department of Mechanical Engineering, Eindhoven University of Technology, P.O.~Box~513, 5600~MB~Eindhoven, The~Netherlands}
\cortext[correspondingauthor]{Corresponding author.}

% ULB address
\address[ULB]{Bulding, Architecture and Town Planning, Universit\'{e} Libre de Bruxelles, 50 Avenue F.D.~Roosevelt, CP~194/2, B--1050 Brusseles, Belgium}

% M2i address
\address[M2i]{Materials Innovation Institute (M2i), P.O.~Box~5008, 2600~GA Delf, The Netherlands}

% TU/e building environment address
\address[TUeB]{Deparment of the Built Environment, Eindhoven University of Technology, P.O. Box~513, 5600~MB Eindhoven, The Netherlands}

%\tnotetext[titlefoot]{The post-print version of this article is published in \emph{J. Mech. Phys. Solids}, \href{https://www.sciencedirect.com/science/article/pii/S0022509618306148}{10.1016/j.jmps.2018.08.019}.}

\begin{abstract}

Materials like paper, consisting of a network of natural fibres, exposed to variations in moisture, undergo changes in geometrical and mechanical properties. This behaviour is particularly important for understanding the hygro-mechanical response of sheets of paper in applications like digital printing. A two-dimensional microstructural model of a fibrous network is therefore developed to upscale the hygro-expansion of individual fibres, through their interaction,  to the resulting overall expansion of the network. The fibres are modelled with rectangular shapes and are assumed to be perfectly bonded where they overlap. For realistic networks the number of bonds is large and the network is geometrically so complex that discretizing it by conventional, geometry-conforming, finite elements is cumbersome. The combination of a level-set and XFEM formalism enables the use of regular, structured grids in order to model the complex microstructural geometry. In this approach, the fibres are described implicitly by a level-set function. In order to represent the fibre boundaries in the fibrous network, an XFEM discretization is used together with a Heaviside enrichment function. Numerical results demonstrate that the proposed approach successfully captures the hygro-expansive properties of the network with fewer degrees of freedom compared to classical FEM, preserving desired accuracy.

\end{abstract}


\end{frontmatter}

%\linenumbers

%
%-----------------------------------------------------------------------------
%	INTRODUCTION
%-----------------------------------------------------------------------------
%
\section{Introduction}
\label{sec:introduction}
\begin{figure}
	\centering
	\subfloat[micrograph of the surface of a printing paper sheet]{\includegraphics[width=60mm, height=60mm]{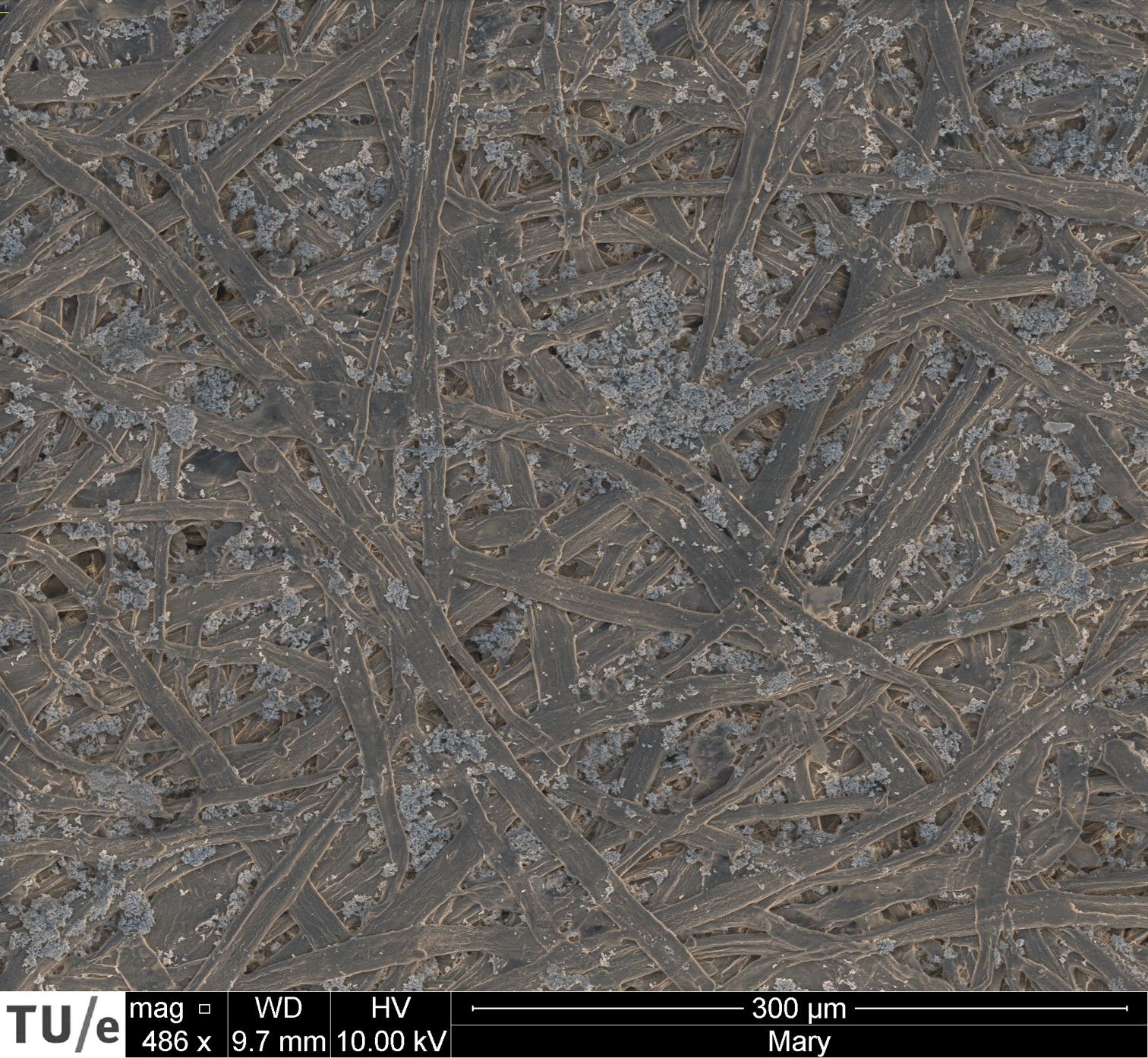}\label{micro}} 
	\hspace{1em}
	\subfloat[an idealized bond between two fibres]{\includegraphics[width=60mm, height=60mm]{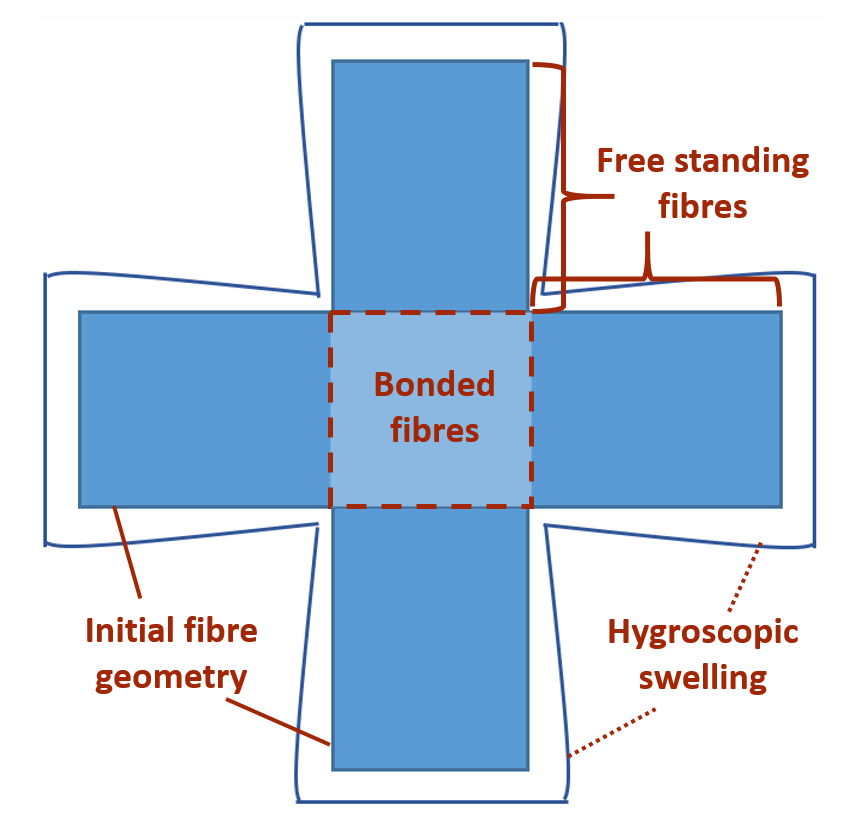}\label{2fibi}} 
	\caption{Illustration of the microstructure of a paper sheet and the role of fibre bonds in hygro-expansion.}
\end{figure}

At the micro-scale, a paper sheet consists of a network of fibres that is produced from wood pulp as shown in Fig.~\ref{micro}. The paper fibres have a preferential orientation (machine direction) due to the manufacturing process,  which results in the observed anisotropic behaviour~\citep{Ada}. In the network, fibres are bonded with each other in certain regions and free standing elsewhere. Upon exposure of a sheet of paper to a humid environment or liquid water, moisture-induced swelling takes place, which is called hygro-expansion~\citep{Larsson}. In practical applications related to printing, this results in macro-scale effects such as curling, waviness, and cockling. At the microstructural level, the changes in each individual fibre due to the variation in moisture content are transmitted to the neighbouring fibres by the bonds in the fibrous network as sketched in Fig.~\ref{2fibi}. These changes include geometrical variations of the shape in each fibre and the accompanying stress concentrations induced by the bonded areas. Understanding these phenomena and their dependence on the properties of the individual fibres and the network geometry is essential for predictive modelling of sheet-scale phenomena in paper as well as other networks of natural fibres.

Gaining insight in the behaviour of a fibrous network subjected to hygroscopic swelling is essential to unravel the influence of different properties of the fibres and network on the overall hygro-expansive response. Among the early attempts to model the mechanical response of paper, \cite{Cox} studied the effect of orientation of fibres on stiffness and strength of paper by assuming that fibres carry only axial forces. The flexural stiffness of the fibres was also taken into account~\citep{Vanden}. The deformation of bonds and the elasto-plastic behaviour of fibres was also studied~\citep{Rama}. Assuming constant strain, the transverse properties of fibres were included~\citep{Schulgasser}. Later, a model incorporating shear forces, axial and bending and torsional moments between rigid bonds was developed for a network of fibres~\citep{Starzewski}. Inter-fibre bonding was also considered  in modelling  the fibrous network~\citep{Bronkhorst}. Some of the recently developed numerical models for fibrous networks mostly modelled the fibrous material with two-dimensional assumptions considering the individual fibres as trusses or beams with only isotropic properties~\citep{Kulachenko,Shah,Dirren}. Most of these works were carried out for the description of the mechanical behaviour of the fibrous network, without any coupling to the hygroscopic response.

Several other studies addressed the hygro-expansivity and dimensional stability of paper. Some studies focused on the relation between hygro-expansion and moisture content of paper in order to understand the factors affecting them~\citep{Nordman,Salmen1,Uesaka2}. Also, studies on the dimensional stability of paper at the macro-scale were carried out~\citep{Page2,Uesaka1,Niskanen1,Niskanen2,Larsson,Erkilla,Bosco1}. Several other works were done on studying the hygro-expansion of paper fibres. However, most of these studies lack in modelling the hygro-mechanical behaviour of complex fibrous networks like paper subjected to moisture fields.

\begin{figure}[t]
	\centering
	\scalebox{1.0}{
	\begin{tikzpicture}[>=stealth]
		\linespread{1}
		\tikzset{
			mynode/.style={inner sep=0,outer sep=0},
			myarrow/.style={myblue,thick},
		}	
		
		% Specimen and mve
		\begin{pgfonlayer}{bg}
		\node[mynode] (specimen) {
			\subfloat[complex fibre network]{\includegraphics[height=60mm]{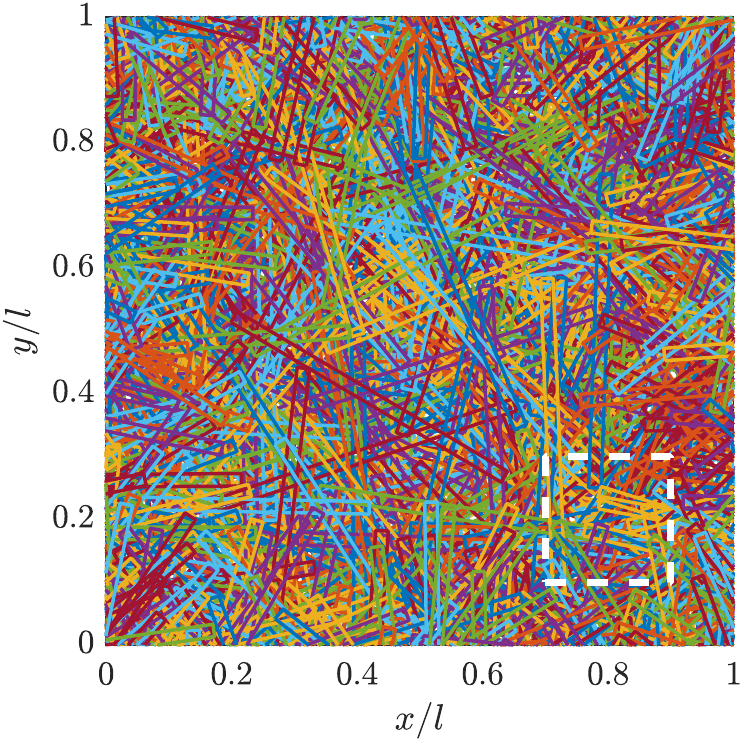}\label{netta}}
		};
		\end{pgfonlayer}
		\node[mynode,right=0.5em of specimen.south east,anchor=south west] (mve) {
			\subfloat[magnified view]{\includegraphics[height=50mm]{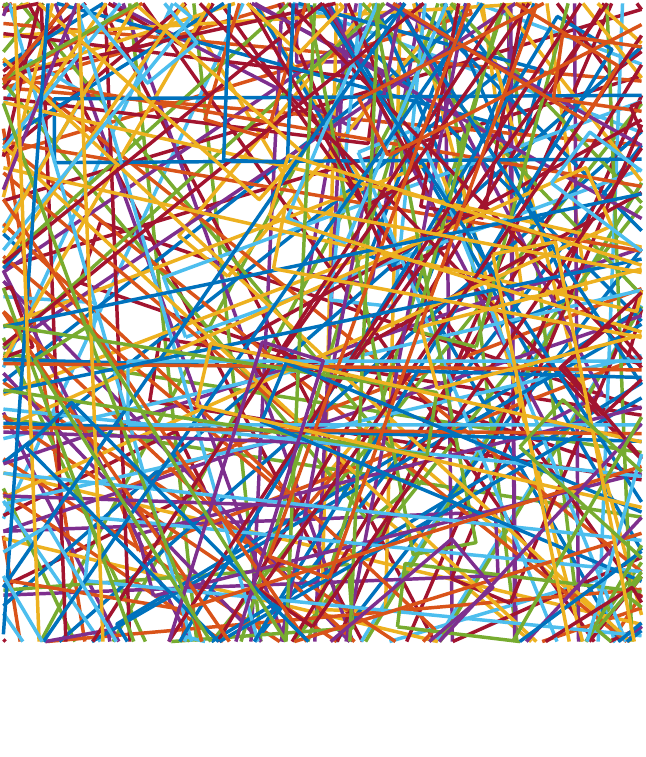}\label{nettb}}
		};

		% Zooming lines in the background layer
		% \begin{pgfonlayer}{bg}
		% \draw[white,dashed,line width=0.5mm] (2.4,-1.6) to (mve.south west);
		% \draw[white,dashed,line width=0.5mm] (1.4,-0.6) to (mve.north west);
		% \end{pgfonlayer}
		
		% Plot auxiliary coordinate grid
		% \draw[help lines,step=0.2] (-5,-2) grid (5,6);
		% \draw[help lines,line width=.6pt,step=1] (-5,-2) grid (5,6);
		% \foreach \x in {-5,-4,-3,-2,-1,0,1,2}
		% \node[anchor=north] at (\x,-2) {\x};
		% \foreach \y in {-2,-1,0,1,2}
		% \node[anchor=east] at (-3,\y) {\y};
		
	\end{tikzpicture}}
	\caption{A model of a typical fibrous network represented as a periodic unit cell, corresponding to coverage~$c = 10$ and the isotropic case ($q = 0$). Each rectangular outline represents a single fibre, where the colour is used to distinguish individual fibres more clearly.}
	\label{nett}
\end{figure}

The aim of the present study is to model the hygro-mechanical behaviour of fibrous materials through a multi-scale analysis of the network using periodic  homogenization~\citep{Guedes,Boutin,Peerlings}. The fibres are modelled as two-dimensional ribbon-like elements in a network subjected to hygro-expansion, for which a finite element framework was used in~\citep{Bosco2}. This study used a regular grid of triangular finite elements to model the network which inevitably is not aligned with most of fibres. The use of such non-conforming finite elements does not allow to accurately capture the geometry of the fibres and of the voids and bonded regions between them. Non-conforming FEM considers any fibre lying inside the centroid of a finite element to contribute to its stiffness. This leads to a geometrical representation of a fibre in the network with jagged boundaries. To better resolve the fibre boundaries, a very fine discretization was therefore employed in~\citep{Bosco2}, which increases the computational effort. A conforming discretization would avoid these limitations. However, in networks of realistic complexity it will be very hard to generate a conforming mesh for all fibres in the densely bonded regions. An example of such a network is shown in Fig.~\ref{netta}, where a paper sheet is represented as a collection of ribbon-shaped rectangular fibres with coverage~$10$, where the coverage is defined as the ratio of the total area occupied by the fibres in the network to the area of the microstructural unit cell. The density of fibre intersections and overlapping regions is more clearly visible in the magnified view in Fig.~\ref{nettb}. The main reasons rendering a conforming triangulation infeasible are:
(i) finding the geometrical intersections of individual fibres and their overlapping regions requires a large number of operations and hence will be too expensive (if possible at all);
(ii) meshing the decomposed geometry will result in an excessively fine triangulation, rendering the subsequent analysis of the network model untractable. 
Hence, there is a need to develop a methodology that captures the fibre boundaries with a better accuracy than a non-conforming mesh and which is less computationally demanding than a conforming discretization strategy.

Here, an advanced discretization scheme is combined with the hygro-mechanical model developed by~\citep{Bosco2} to enable modelling of larger systems in two-dimensional configurations, with the potential to be extended towards three dimensions. The level-set formalism is used in combination with the Extended Finite Element Method (XFEM) to capture the geometrical complexity of the network---here in two dimensions---in an efficient and inexpensive manner, providing the decomposed geometry and mutual intersections of individual fibres. The level-set methodology is a mathematical tool to describe complex and time-evolving boundaries (or, more generally, geometry) implicitly~\citep{Sethian}. The boundary of a fibre is represented by the zero level-set of a higher dimensional function. This results in a versatile geometrical description that is de-coupled from the spatial discretization. XFEM~\citep{Daux} allows to account for the effect of interfaces (boundaries of fibres) on the mechanical behaviour of the problem. In the bonded regions, the interpolation functions classically used in the  elements are modified by discontinuous enrichment functions, so that geometrical discontinuities associated with the fibre interfaces can be resolved. This allows to capture the fibre's boundaries in the bonded regions accurately using reasonably coarse meshes, by coupling of the level-set functions with the XFEM enrichment. For realistic networks, which are as complex as the one shown in Fig.~\ref{nett}, the level set technology thus provides a feasible numerical tool for the description of their geometry.

This paper is organized as follows: In Section~\ref{sec:fibre_network}, the geometry of the fibre network model is discussed briefly, along with the level-set formalism used to represent it. In Section~\ref{sec:discretization_strategy}, the XFEM discretization used for the fibrous network is described, together with the required specific numerical integration scheme to accommodate it. Section~\ref{sec:results} presents the simulation results of the hygro-mechanical behaviour of paper using the XFEM enrichment, first for the simplified models presented in~\citep{Bosco1}, next for somewhat more complex networks with different coverages, and finally for a complex realistic network. The local and global responses are analysed for both medium-complexity and complex networks. Finally, Section~\ref{sec:conclusions} reports the conclusions and perspectives.

Throughout this contribution, the following notations are used for operations on the Cartesian tensors. Scalars, vectors, and tensors are denoted by~$a$, $\vec{a}$, and $\boldsymbol{A}$ respectively. The $4$-th order tensors are represented by~$\tensorfour{A}$. For tensor and vector operations, the following equivalent notations are used with Einstein's summation convention on repeated indices: $\boldsymbol{A}:\boldsymbol{B} = A_{ij}B_{ji}$ with ($i=x,y,z$ for the global reference system and $i=l,t,z$ for the local reference system). The Voigt notation used to represent tensors and tensor operations in a matrix format is depicted as follows: $\column{a}$ and $\underline{A}$ denote a column matrix and a matrix of scalars respectively. The matrix multiplication is denoted as~$\underline{A} \, \underline{a} = A_{ij}a_j$.
%
%-----------------------------------------------------------------------------
%	FIBRE NETWORK MODEL
%-----------------------------------------------------------------------------
%
\section{Fibre network model}
\label{sec:fibre_network}
%
%----------------------------------
%	FIBRE MODEL
%----------------------------------
%
\subsection{Fibre model}
\label{sec:fibre_model}
\begin{figure}
	\centering
	\includegraphics[width=55mm, height=50mm]{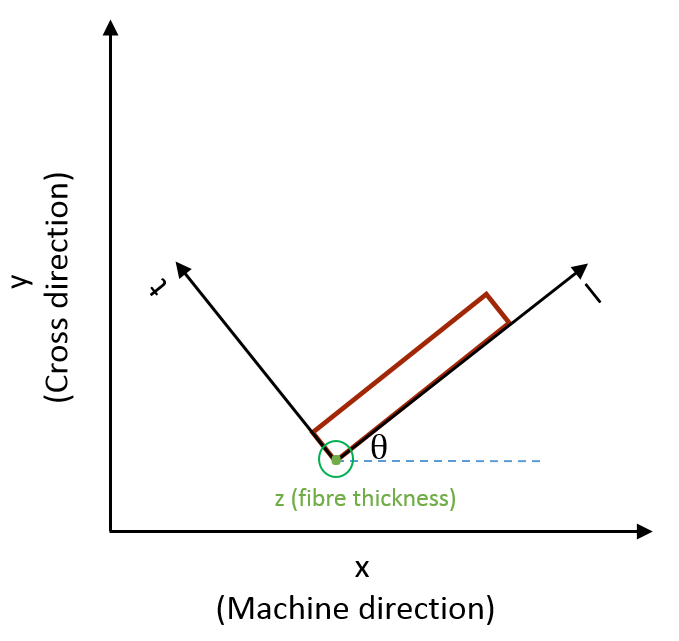}
	\caption{The local and global coordinate axes.}
	\label{axi}
\end{figure}

The fibre-level constitutive model formulated in~\citep{Bosco1} is adopted here. A two-dimensional plane stress model is assumed and the fibres in consideration are oriented in the~$(x,y)$ plane of the~$(x,y,z)$ global reference frame, and subjected to a uniform moisture change. For the fibre constitutive model, a local reference frame~$(l,t,z)$ is considered along the  directions of the fibre as shown in Fig.~\ref{axi}.
The hygro-mechanical properties of the paper fibres are assumed to be transversely isotropic with respect to their longitudinal axis. The general elastic constitutive law for a fibre, assuming plane stress state in the $z$-direction, and  subjected to a moisture change~$\Delta\chi$ is expressed as
\begin{equation}
\boldsymbol{\sigma}^f = {^4\boldsymbol{D}^f}:({\boldsymbol\varepsilon^f}-{^h\boldsymbol\varepsilon^f}),
\label{const}
\end{equation}
where the hygroscopic strain~${^h}\boldsymbol\varepsilon^f$ is given by
\begin{equation}
{^h}\boldsymbol\varepsilon^f = \boldsymbol\beta^f\Delta\chi.
\label{epsih1}
\end{equation}
In these expressions, ${^4}\boldsymbol{D}^f$, $\boldsymbol\varepsilon^f$, and~$\boldsymbol\beta^f$ are the elastic constitutive tensor, the strain tensor and the hygro-expansivity tensor of the fibre respectively. In matrix notation, ${^4}\boldsymbol{D}^f$ and~$\boldsymbol\beta^f$ are represented as
\begin{equation}
\underline{D}^f=
\begin{pmatrix}
\frac{E_l}{(1-\nu_{lt}\nu_{tl})} & \frac{E_l\nu_{tl}}{(1-\nu_{lt}\nu_{tl})} & 0 \\
\frac{E_t\nu_{lt}}{(1-\nu_{lt}\nu_{tl})} &
\frac{E_t}{(1-\nu_{lt}\nu_{tl})} & 0 \\ 
0 & 0 & G_{lt}
\end{pmatrix}
, \quad
\underline{\beta}^f=
\begin{pmatrix}
\beta_l\\
\beta_t\\
0
\end{pmatrix}.
\label{mat}
\end{equation}
In Eq.~\eqref{mat}, $E_l$ and~$E_t$ denote the elastic moduli in the longitudinal and transverse direction with respect to the fibre axis, $G_{lt}$ is the in-plane shear modulus, and~$\nu_{lt}$ and~$\nu_{tl}$ in-plane Poisson ratios. The coefficients of hygroscopic expansion, $\beta_l$ and~$\beta_t$, are different in the longitudinal and transverse directions of the fibre.

In the local reference system, a fibre~$i$ is oriented at an angle~$\theta_i$ with respect to the global reference system~$(x,y,z)$. Therefore, the relationships of Eqs.~\eqref{const}--\eqref{mat} need to be transformed from the local (fibre) frame~$(l,t,z)$ to the global frame of the paper material~$(x,y,z)$ for each fibre in the network~\citep{Roylance}. The fibre bonds are important because of their influence on the overall behaviour of the network. In the 2D modelling, the fibres are assumed to be perfectly  bonded to simulate the interplay between hygro-expansion and elasticity of the fibres. This implies full displacement strain compatibility inside a bond for all the fibres interconnected in that bond.
%
%----------------------------------
%	LEVEL-SET FORMALISM
%----------------------------------
%
\subsection{Level-set formalism}
\label{sec:level_set_formalism}
\begin{figure}[htbp]
	\centering
	\includegraphics[clip,trim={0mm 0mm 0mm 0mm},width=60mm, height=25mm]{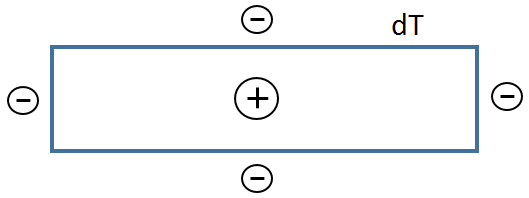}
	\caption{Sign convention used for signed distance function representing the rectangular fibres.} 
	\label{sigrect}
\end{figure}

The level-set formalism~\citep{Sethian,Fedkiw} is here used to describe the geometry of fibres. In this method, the fibre is described as the zero level-set of a higher dimensional level-set function~$\phi(\vec{x})$ (see Fig.~\ref{sigrect}).
In most cases, including the current work, the level-set function gives the signed distance of a point~$\vec{{x}}$ to the interface of the rectangular fibre denoted by~$dT$ in Fig.~\ref{sigrect}. The level-set function~$\phi(\vec{{x}})$ is defined as
\begin{equation}
\phi(\vec{{x}}) = \pm \underset{ \vec{{x}}_T \in dT }{ \min } \| \vec{x}-\vec{x}_T \|,
\end{equation}
where the sign is negative if~$\vec{{x}}$ is outside and positive if it is inside the contour defined by~$dT$.
%
%----------------------------------
%	RANDOM FIBRE NETWORK CREATION
%----------------------------------
%
\subsection{Random fibre network creation}
\label{sec:random_fibre_network_creation}
\begin{figure}
	\centering
	\subfloat[fibre orientation distribution function]{\includegraphics[trim={8mm 3mm 0mm 0mm},width=70mm,height=63.5mm]{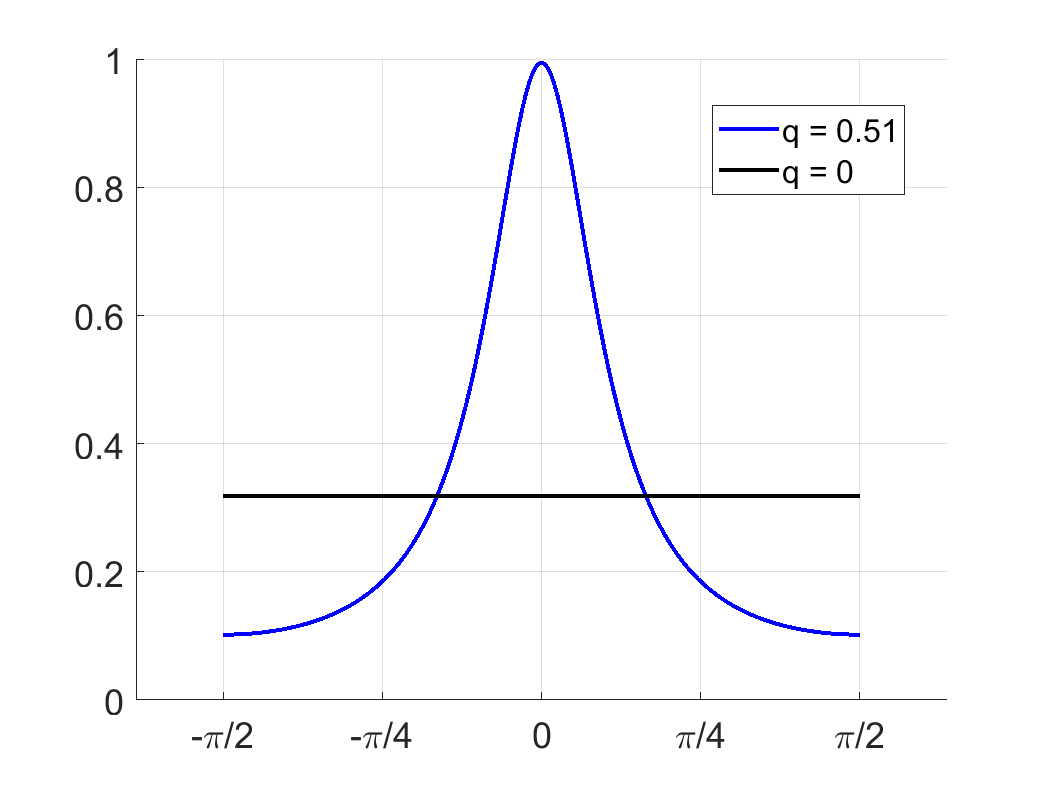}\label{ani}}
	\hspace{0.1em}
	\subfloat[periodic unit cell]{\includegraphics[trim={0mm 0mm 0mm 0mm},width=70mm, height=60mm]{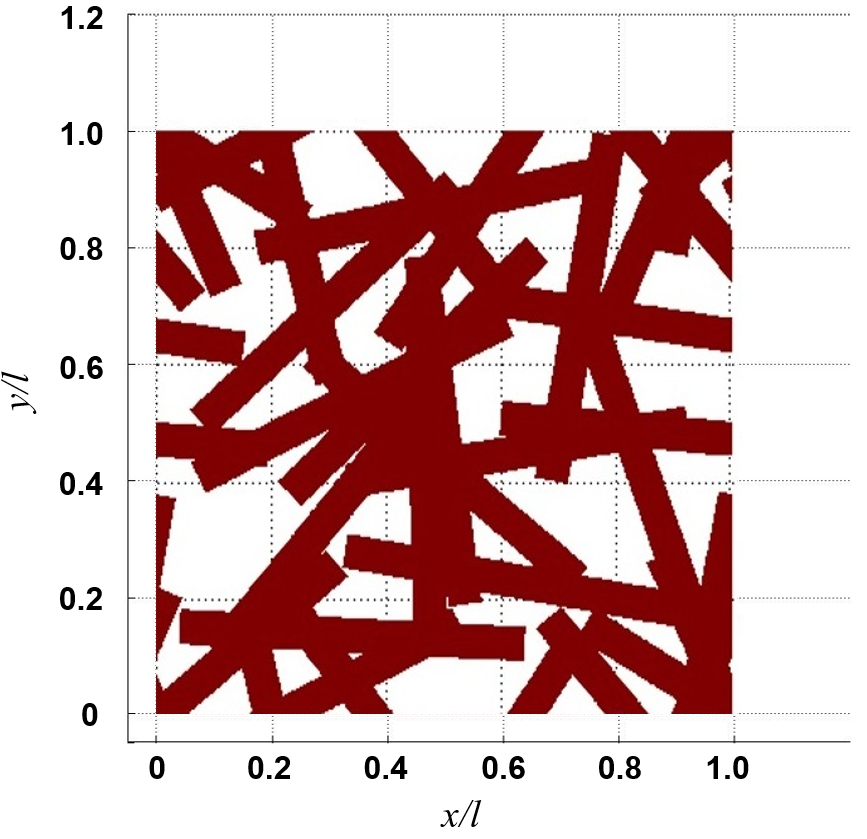}\label{fibii}}
	\caption{Anisotropic orientation function and  a random network of fibres.} 
	\label{fig:orientation}
\end{figure}

In order to understand the hygroscopic behaviour of a complex network of fibres, a set of rectangular fibres having a length~$l_f$ and width~$w_f = l_f/10$ is randomly generated in a unit cell of length~$l = 5l_f/3$. Each of the fibres is generated with random coordinates for its centroid ~$[x,y] \in [0,l]$ as shown in Fig.~\ref{fibii}. Now, depending on the number of fibres~$n$ generated in the unit cell, a coverage~$c$ is defined as the ratio of the total area occupied by all fibres and the area of the unit cell.
The orientations of the randomly generated fibres in the network satisfy a probability density function based on~\citep{Cox}:
\begin{equation}
f(\theta) = \frac{1}{\pi}\frac{ 1-q^2 }{ 1+q^2-2q\cos{(2\theta)} },
\label{eq:distribution}
\end{equation}
where~$\theta$ is the angle between the fibre and the machine direction, with~$-\pi/2 < \theta < \pi/2$. Fig.~\ref{ani} shows the probability density function for the orientation for two values of the anisotropy parameter~$q$: for the isotropic case~$q=0$, and an arbitrary degree of anisotropy~$q=0.51$ \citep{Bosco1}.

Once the fibres are generated and each of them is assigned a level-set function, $\phi_i(\vec{x})$, $i = 1,\dots,n$, a periodicity condition is enforced on the unit cell. The level-set  function that allows identifying the fibres and voids for the entire network can be described as 
\begin{equation}
\psi{(\vec{x})} = \max(\phi_1(\vec{x}),\phi_2(\vec{x}),\dots,\phi_n(\vec{x})),
\end{equation}
where~$\psi(\vec{x}) \geq 0$ in fibres and~$\psi(\vec{x}) < 0$ in voids.
%
%-----------------------------------------------------------------------------
%	DISCRETIZATION STRATEGY
%-----------------------------------------------------------------------------
%
\section{Discretization strategy}
\label{sec:discretization_strategy}
%
%----------------------------------
%	XFEM METHODOLOGY
%----------------------------------
%
\subsection{XFEM methodology}
\label{sec:xfem_methodology}
The extended finite element method~(XFEM) is a numerical discretization method developed to model material discontinuities, singularities, and moving boundaries independently of the underlying finite element mesh. With the microstructures considered in this work, it is not feasible to produce a geometry-conforming mesh for the free standing parts of fibres and even more in the bonded regions where fibres partially overlap. Accurate local results cannot be obtained by the conventional finite element method (FEM) if the  mesh does not conform to the boundaries of the fibres. XFEM can be used instead to capture the fibre boundaries both in free standing and bonded regions accurately even with a regular non-conforming mesh. In terms of accuracy, an adequate solution can be obtained by enriching the classical shape functions with special functions capturing the geometrical discontinuities in XFEM.

Depending on the nature of the problem under consideration, a suitable enrichment function needs to be adopted in XFEM. To model the geometrical discontinuity between solid (fibres) and voids in a fibrous network with a regular mesh, a Heaviside function defined in terms of the level-set function~$\phi$ is adopted~\citep{Sukumar} as an enrichment function, since the geometry is already described by the level-set function~$\psi$. The displacement interpolation then reads~\citep[][Section~4]{Daux}
\begin{equation}
\vec{u}(\vec{x}) = \sum_{e = 1}^m H(\psi(\vec{x})) N_e(\vec{x}) \underline{u}_e,
\label{enrich11}
\end{equation}
in which~$m$ is the number of all elements of the underlying mesh and~$\underline{u}_e$ is the nodal displacement column of the finite element~$e$, $H(\psi(\vec{x})) = 1$ for~$\psi(\vec{x}) \geq 0$ inside fibres and~$ H(\psi(\vec{x})) = 0 $ for~$ \psi(\vec{x}) < 0$ inside void regions.

The interconnection between XFEM (to model discontinuities) and level-set (to capture the geometry), referred to as LS-XFEM in what follows, is discussed in detail in the next section. A LS-XFEM formalism that makes use of the level-set information in this particular XFEM setting thus greatly simplifies the set up of the model under consideration.
%
%----------------------------------
%	DISCRETIZATION WITH THE LEVEL-SET AND XFEM FORMALISM (LS-XFEM)
%----------------------------------
%
\subsection{Discretization with the level-set and XFEM formalism (LS-XFEM)}
\label{sec:discretization}
In the current work, the behaviour of the fibres is assumed elastic and small strains and displacements are considered. The total potential energy in the unit cell consisting of network of fibres is given by
\begin{equation}
\pi = \frac{1}{2}\int_{V} \sum_{i=1}^{n} H(\phi_i(\vec{x})) \, ( \bs{\varepsilon}^f - {^h}\bs{\varepsilon}^f_i ) : \bs{D}_i^f : ( \bs{\varepsilon}^f - {^h}\bs{\varepsilon}^f_i ) \, t_i \, \mathrm{d}A,
\end{equation}
where~$n$ is the total number of fibres occupying the periodic cell area~$A$, $\phi_i(\vec{x})$ is the signed distance function of the~$i$-th fibre, $\bs{\varepsilon}^f$ the strain tensor associated with Eq.~\eqref{enrich11}, ${^h}\bs\varepsilon^f_i$ the hygroscopic strain in fibre~$i$, and~$t_i$ its thickness. This expression may be written in matrix format as
\begin{equation}
\begin{aligned}
\pi
&= \int_{V}\sum_{i=1}^{n} H(\phi_i(\vec{x})) \frac{1}{2} (\underline{\varepsilon}^f - {^h}\underline{\varepsilon}^f_i)^T \underline{D}_{i}^f(\underline{\varepsilon}^f - {^h}\underline{\varepsilon}^f_i) \, \mathrm{d}V \\
&=
\int_{V} \sum_{i=1}^{n} H(\phi_i(\vec{x})) \frac{1}{2} \left[ (\underline{\varepsilon}^f)^T \underline{D}_{i}^f \underline{\varepsilon}^f - 2(\underline{\varepsilon}^f )^T \underline{D}_{i}^f {^h}\underline{\varepsilon}^f_i + ({^h}\underline{\varepsilon}^f_i) \underline{D}_{i}^f{^h}\underline{\varepsilon}^f_i \right] \mathrm{d}V,
\end{aligned}
\label{inte11}
\end{equation}
in which~$\underline{D}^{f}_i$ and~${^h}\underline{\varepsilon}^f_i$ are the elastic constitutive matrix and hygroscopic strain matrix transformed to the global frame, for a particular fibre~$i$ aligned at an angle~$\theta_i$ with respect to the global frame as discussed earlier. Using the strain-displacement relationship for each finite element, ${\underline{\varepsilon}}^f = \underline{{B}}_e \underline{u}_e$, where~$\underline{B}_e$ is the standard strain-displacement matrix, the integration over the entire network of fibres in Eq.~\eqref{inte11} is carried out element wise. Swapping the summations over all~$m$ finite elements with areas~$A_e$ and the~$n$ fibres, the assembled potential energy is obtained as
\begin{equation}
\begin{aligned}
\pi
&=
\frac{1}{2} \sum_{e=1}^{m} \left[ \underline{u}_e^T \int_{A_e}\sum_{i=1}^{n} H(\phi_i(\vec{x})) \, \underline{B}_e^T \underline{D}_i^f \underline{B}_e \, t_i \, \mathrm{d}A \, \underline{{u}}_e \right. \\ 
&
\left. - 2\int_{A_e} \underline{u}_e^T \sum_{i=1}^{n} H(\phi_i(\vec{x})) \, \underline{B}_e^T \underline{D}_{i}^f {^h}\underline{\varepsilon}_i^f \, t_i \, \mathrm{d}A
+
\int_{A_e} \sum_{i=1}^n H(\phi_i(\vec{x})) {^h}\underline{\varepsilon}^f_i \underline{D}_{i}^f {^h}\underline{\varepsilon}_i^f \, t_i \, \mathrm{d}A \right].
\end{aligned}
\label{allf}
\end{equation}
Of all the possible displacements that satisfy the boundary conditions of an elastic structural system, the ones corresponding to the equilibrium configuration minimize the total potential energy, i.e.
\begin{equation}
\frac{\partial\pi}{\partial\underline{{u}}} = \underline{0},
\end{equation}
or, using Eq.~\eqref{allf},
\begin{equation}
\sum_{e=1}^m \sum_{i=1}^n \int_{A_e} H(\phi_i(\vec{x})) \, \underline{B}_e^T \underline{D}_i^f \underline{B}_e\, t_i \, \mathrm{d}A \, \underline{{u}}_e 
-
\sum_{e=1}^m \sum_{i=1}^n \int_{A_e} H(\phi_i(\vec{x})) \, \underline{B}_e^T \underline{D}_i^f~^h\underline{\varepsilon}_i^f \,  t_i \, \mathrm{d}A = \underline{0}.
\end{equation}
This equation is of the form of a linear system
\begin{equation}
\underline{K} \, \underline{u} - \underline{f} = \underline{0},
\end{equation}
in which~$\underline{u}$ collects all nodal displacements, and stiffness matrix and the hygroscopic load vector are given by
\begin{align}
\underline{K} &= \sum_{e=1}^m \sum_{i=1}^n \int_{A_e} H(\phi_i(\vec{x})) \, \underline{B}_e^T \underline{D}_i^f \underline{B}_e \, t_i \, \mathrm{d}A,\label{stifff} \\
\underline{f} &= \sum_{e=1}^m \sum_{i=1}^n \int_{A_e} H(\phi_i(\vec{x})) \, \underline{B}_e^T \underline{D}_i^f {^h}\underline{\varepsilon}_i^f \, t_i \, \mathrm{d}A.\label{forcevec}
\end{align}
Linear triangular finite elements are used for the discretization. Three cases can be distinguished for mapping fibres onto a finite element. First, if the finite element lies entirely in a void, the Heaviside function in Eqs.~\eqref{stifff}--\eqref{forcevec} vanishes and the element does not contribute to the stiffness matrix and hygroscopic load vector. If the finite element lies entirely in one or more fibres, its full contribution is accounted for in the computation of the stiffness matrix and hygroscopic load vector. Finally, if the finite element is intersected by fibre boundaries, a specific integration technique needs to be used, as described next.
%
%----------------------------------
%	NUMERICAL SOLUTION SCHEME
%----------------------------------
%
\subsection{Numerical integration scheme}
\label{eq:numerical_integration}
\begin{figure}
	\centering
	\subfloat[]{\includegraphics[scale=0.45]{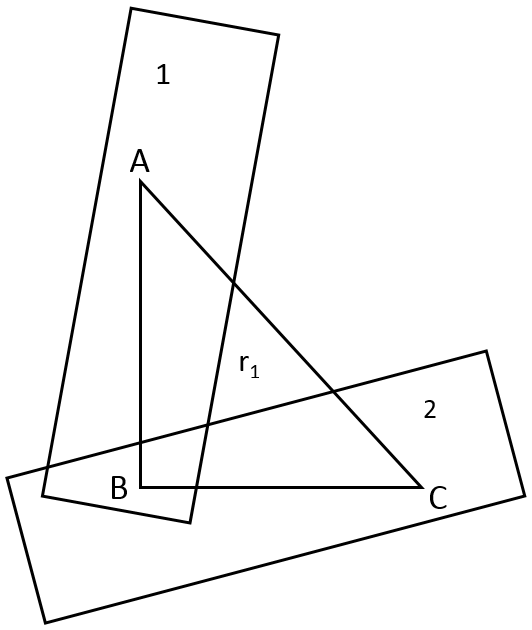}\label{trian11a}}
	\hspace{0.5em}
	\subfloat[]{\includegraphics[scale=0.45]{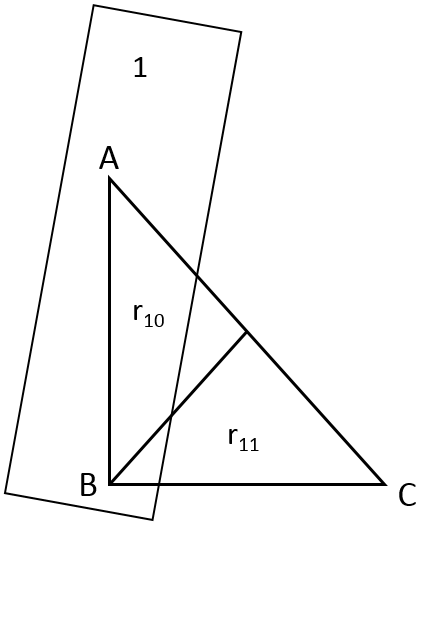}\label{trian11b}}
	\hspace{0.5em}	
	\subfloat[]{\includegraphics[scale=0.45]{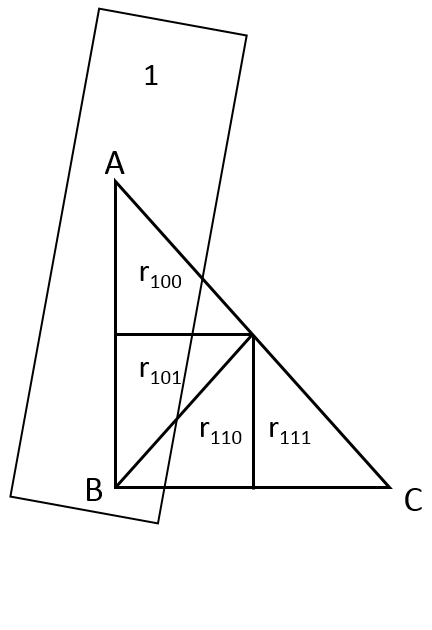}\label{trian11c}}
	\hspace{0.5em}	
	\subfloat[]{\includegraphics[scale=0.45]{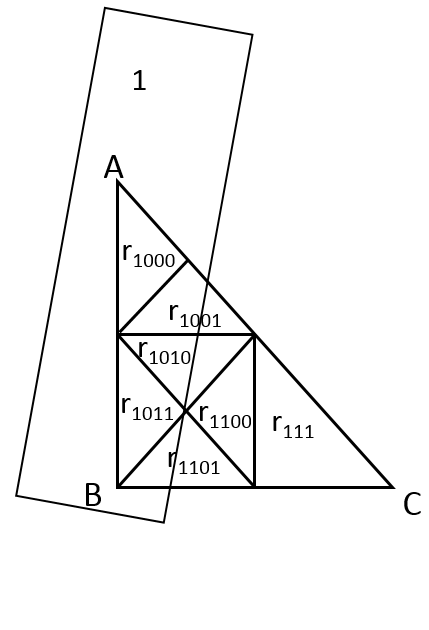}\label{trian11d}}
	\caption{Partitioning of a triangular finite element intersected by the boundaries of two (rectangular fibres) for the area integration. (a)~The finite element~$r_1$ is partially covered by two fibres. (b)~As a first step, fibre~$1$ is considered. Since the finite element lies partially in the fibre (the level-set values at the vertices are not all of the same sign), it is bisected along the longest edge, resulting in triangles~$r_{10}$ and~$r_{11}$. (c)~Both of the newly formed triangles~$r_{10}$ and~$r_{11}$ are again lying partially in fibre~$1$. They are hence further bisected. (d)~Triangles~$r_{100}$, $r_{101}$, and~$r_{110}$ are further split; $r_{111}$ remains unaffected as it is completely outside the fibre. The partitioning further continues until a certain tolerance is reached.}
	\label{trian11}
\end{figure}

In this section, the integration scheme used for elements intersected by fibre boundaries is detailed. The integration over a particular finite element for all fibres passing in it is considered for the stiffness and force vector of that element, recall Eqs.~\eqref{stiffsim} and~\eqref{forcesim},
\begin{align}
\underline{K}_e &= \sum_{i=1}^n \underline{B}_e^T \underline{D}_i^f \underline{B}_e \int_{A_e} H(\phi_i(\vec{x})) \, t_i \, \mathrm{d}A,\label{stiffsim}\\
\underline{f}_e &= \sum_{i=1}^n \underline{B}_e^T \underline{D}_i^f {^h}\underline{\varepsilon}_i^f \int_{A_e} H(\phi_i(\vec{x})) \, t_i \, \mathrm{d}A.\label{forcesim}
\end{align}
In the computation of the contribution of each fibre to the finite element, it is therefore only required to determine the area occupied by each fibre, i.e.~$A_1, \dots, A_n$ in this particular element, where~$A_i = \int_{A_e} H(\phi_i(\vec{x})) \, \mathrm{d}A$. The algorithm used to compute this area of each fibre in the finite element is illustrated by considering two fibres lying in the region of a triangular element with vertices~$(A, B, C)$ as shown in Fig.~\ref{trian11}. Using the level-set function for a particular fibre~$i$, if one of the vertices of the finite element is identified to lie inside the fibre, then the element is considered to lie in the fibre. Therefore, the triangle is sub-divided  along its longest edge. The subtriangulation continues and stops only when a tolerance is met in terms of the smallest triangle area. This procedure is executed for all fibres lying partially inside the finite element. The integration scheme can be summarized as follows:
\begin{enumerate}
	\item Check for the first fibre if all element vertices lie in it (i.e.~$\phi_1(A) \geq 0, \phi_1(B) \geq 0, \phi_1(C) \geq 0$). If the particular fibre lies partially in the element, then the element is split into two new triangles as shown in Fig.~\ref{trian11b}.
	
	\item The subdivision continues for triangles intersected by the fibre's boundaries Figs.~\ref{trian11c}--\ref{trian11d} If a triangle lies completely inside or outside the fibre, the subdivision is not performed.
	
	\item The sub-triangulation is terminated when a specified tolerance limit in terms of a minimum triangle area is achieved.
	
	\item After the termination of subtriangulation, the area occupied by the fibre in the finite element is sum of the areas of triangles lying completely in the fibre and the ones whose centroid lies inside the fibre. Thereafter, the above steps are repeated again for the next fibre.
\end{enumerate}

The element sub-triangulation for numerical integration is adopted because of its ability to control the integration accuracy in relation to the computational cost. The main reason here is that a conforming integration sub-triangulation for all fibres might become relatively expensive, and for complex systems as the one shown in Fig.~\ref{nett} it may reach a complexity that is comparable to creating a fully conforming triangulation.
%
%----------------------------------
%	MESH REFINEMENT
%----------------------------------
%
\subsection{Mesh refinement}
\label{sec:mesh_refinement}
Considering a unit cell consisting of a network of fibres with complex topology, a coarse triangulation is naturally insufficient for an accurate description of the interfaces. Using a globally fine triangulation would however dramatically increase the computational cost. Therefore, a refinement strategy based on the backward longest edge bisection algorithm~\citep{Rivara} is adopted here for elements intersected by boundaries. This approach preserves adequate accuracy at a lower cost, still allowing for large triangles in regions with little strain fluctuation within fibres and voids. This mesh refinement should not be confused with the subtriangulation discussed in the previous section. The subtriangulation is used to accurately capture the geometry, i.e. the volume of an element occupied by a particular fibre. Mesh refinement, on the other hand, improves the accuracy of the kinematics of the problem.

After the generation of the geometry of a network of fibres, the finite elements located at the boundaries of the fibres are revisited for mesh refinement. For each such element, the corresponding fibre signed distance functions are evaluated at all of its vertices of considered triangular finite element. When all three vertices are contained inside all fibres passing through it, or when none of them belong to the fibres, the element is not categorized as a boundary element. In all other cases, the element is identified as boundary element to be refined, for which a mesh refinement algorithm is applied.

\begin{figure}
	\centering
	\subfloat[]{\includegraphics[scale=0.6]{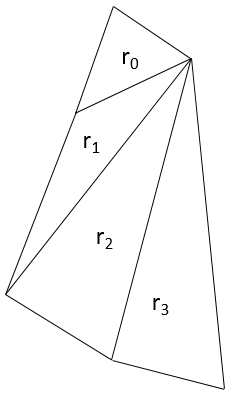}}
	\hspace{0.5em}
	\subfloat[]{\includegraphics[scale=0.6]{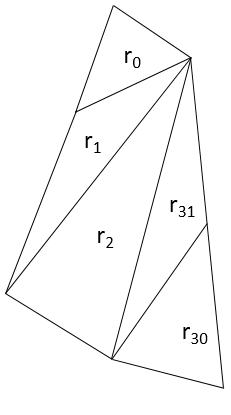}}
	\hspace{0.5em}	
	\subfloat[]{\includegraphics[scale=0.6]{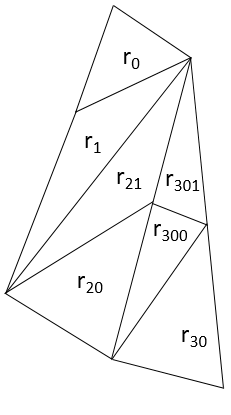}}
	\hspace{0.5em}	
	\subfloat[]{\includegraphics[scale=0.6]{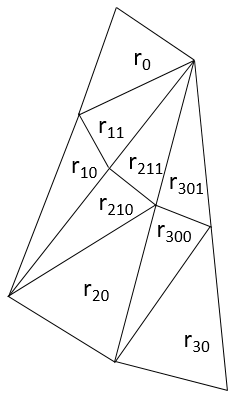}}
	\hspace{0.5em}	
	\subfloat[]{\includegraphics[scale=0.6]{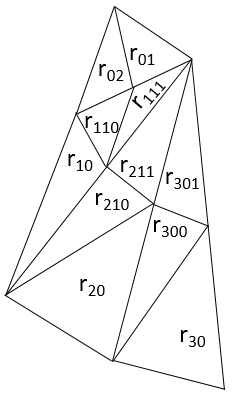}}				
	\caption{Illustration of the longest edge propagation path (LEPP) refinement in a triangular mesh. (a)~The LEPP is given by~$r_0, r_1, r_2, r_3 $ with~$r_3$ being the terminal triangle. (b)~$r_3$ is bisected along its longest edge and the new LEPP is~$r_0, r_1, r_2, r_{31} $, (c)~$r_{31}$ and~$r_2$, being the terminal triangles, are bisected along their longest edge. The LEPP is~$r_0, r_1, r_{21}$. (d)~$r_{21}$ and~$r_1$ are bisected along their longest edge. The LEPP is~$r_0$ and~$r_{11}$. (e)~$r_{11}$ and~$r_0$ are bisected. This concludes the refinement.}
	\label{riva}
\end{figure}

As defined by~\cite{Rivara}, for a triangle~$r_0$ of any triangulation~$T$, the longest edge propagation path (LEPP) is an ordered list of triangles~$r_0, r_1, r_2,\dots,r_n$, such that~$r_i$ is neighbour to~$r_{i-1}$ along its longest edge. The (LEPP) terminates with (a) one triangle with its longest edge along the external boundary of the mesh, or (b) a pair of triangles sharing the same longest edge. Now, based on the LEPP of triangle~$r_0$ marked for refinement, a backward longest edge refinement algorithm is used in which the longest edge of~$r_n$ is bisected in the former case and  both triangles~$r_{n-1}$ and~$r_n$ are bisected along longest edge for the later case. The LEPP is updated and the procedure is repeated until the initial triangle~$r_0$ is bisected as well. In Fig.~\ref{riva}, the longest edge bisection algorithm is illustrated for a simple triangulation.

\begin{figure}[htbp]
	\centering
	\subfloat[refined mesh]{\includegraphics[height=60mm]{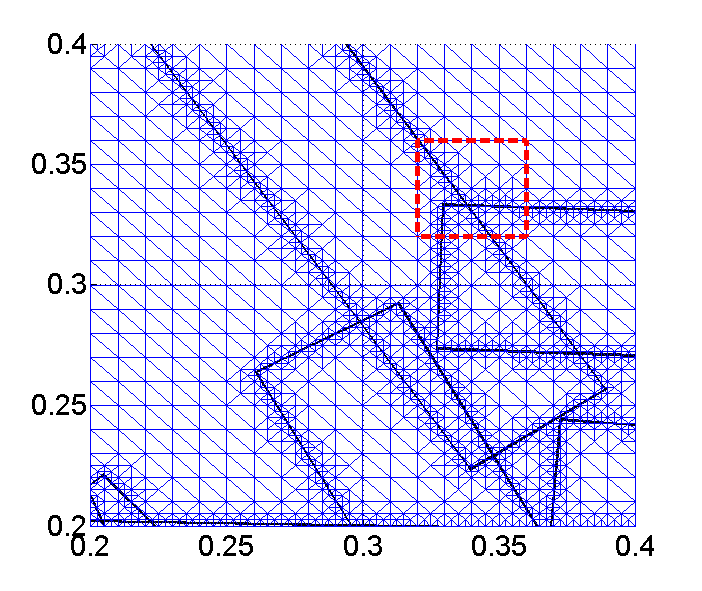}}
	\hspace{0.1em}
	\subfloat[magnified view]{\includegraphics[clip,height=60mm]{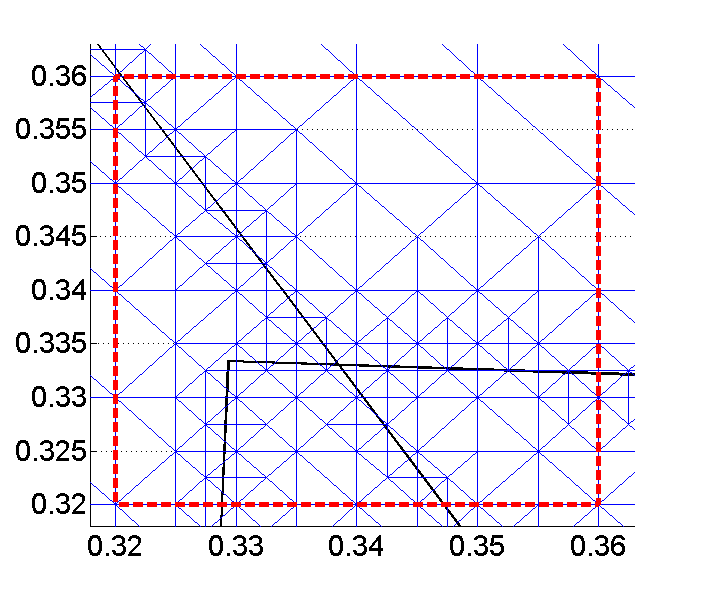}}
	\caption{(a)~A refined mesh. (b)~Magnified view depicting the non-conforming character of the mesh relative to fibre interfaces.}
	\label{meshme}
\end{figure}

In the current work, the elements at the interfaces between a free standing fibre segment and void, as well as at the boundaries of bonds are refined. The newly generated refined triangular elements at these locations also better capture multiple fibres passing through the finite elements. Therefore, the accuracy of the mesh improves in representing the fibre geometry and, more importantly, its kinematics in these complex networks. In addition, the boundary edges of the unit which is periodic in nature are also refined to maintain periodicity of the nodes.

However, the refined mesh, as seen in Fig.~\ref{meshme} will not be conforming to the fibre boundaries. As mentioned earlier, the high number of fibre interfaces particularly in the bonds, would render it difficult to capture with a conforming mesh at cheaper computational effort. This again emphasizes the motivation to use the proposed LS-XFEM formalism, which allows to decouple the mesh and fibre boundaries with a structured mesh that still captures the fibre edges adequately.
%
%-----------------------------------------------------------------------------
%	RESULTS AND DISCUSSION
%-----------------------------------------------------------------------------
%
\section{Results and discussion}
\label{sec:results}
To demonstrate the benefits of the LS-XFEM formalism for fibrous networks, simplified illustrations are first presented below. Subsequently, the formalism is used for medium-complexity and complex realistic networks to study the effect of microstructural features on the sheet-scale properties due to moisture infiltration.
%
%----------------------------------
%	SIMPLIFIED NETWORKS
%----------------------------------
%
\subsection{Simplified networks}
\label{sec:simplified_networks}
First, the stress level in a single family of parallel fibres subjected to a tensile load is assessed, followed by the study of the stress concentrations in the bond regions of an elementary network consisting of two families of fibres subjected to a macroscopic tensile load.
%
%----------------------------------
%	PARALLEL FIBRES SUBJECTED TO UNIAXIAL TENSION
%----------------------------------
%
\subsubsection{Parallel fibres subjected to uniaxial tension}
\begin{figure}
	\centering
	\subfloat[infinitely long fibres and loading considered]{\includegraphics[height=60mm]{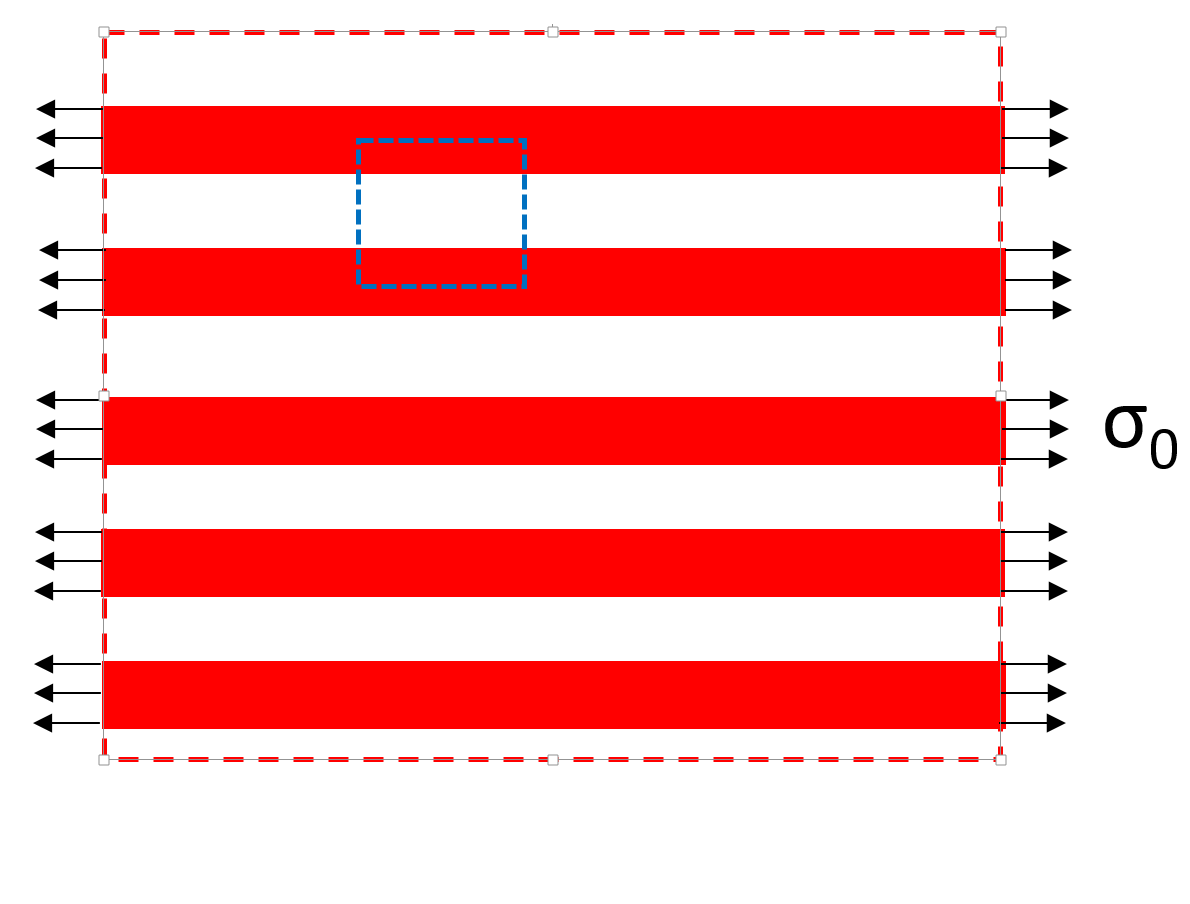}\label{figg2}}
	\hspace{0.5em}	
	\subfloat[discretized unit cell with $20 \times 20 \times 2 = 800$ elements non-conforming to the geometry]{\includegraphics[height=60mm]{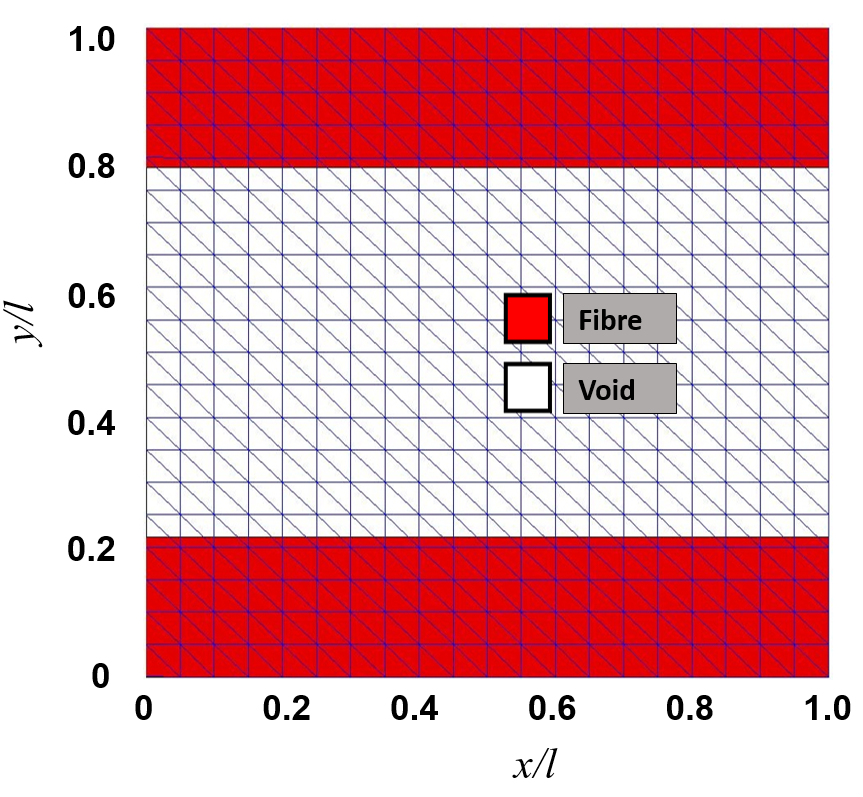}\label{figg1}}
	\caption{(a)~An infinite number of parallel, equispaced, infinitely long fibres subjected to a uniaxial stress~$\sigma_0$. (b)~Corresponding periodic unit cell with a regular mesh non-conforming to the geometry.} 
	\label{pics:infinite_fibres}
\end{figure}

In this problem, an infinite number of parallel, equispaced, infinitely long fibres is considered, as shown in Fig.~\ref{figg2}. The fibres are subjected to a horizontal stress, $\sigma_0$. Using a square periodic unit cell of length~$l$ for this simple geometry, half of a fibre occurs at the top and the another half at the bottom of the cell, see Fig~\ref{figg1}. The chief reason to use this unit cell in the example is that there are regions with fibre and voids, which may also be the case in the complex fibrous network considered later. Also, in the discretized unit cell, the mesh is not conforming to the geometry of the fibre, i.e.~the same situation occurs in a complex fibrous network, and therefore we will see the advantages of the LS-XFEM formalism as compared to the non-conforming FEM. The main reason for using a non-conforming mesh for standard FEM is to measure the error induced by it. Although a mesh conforming to the considered geometry could easily be obtained for this simple system, it is not possible for much more complex systems, as discussed in Section~\ref{sec:realistic_network} below. The input parameters for the anisotropic fibres are an elastic modulus in the longitudinal direction, $E_l$, and in the transverse direction, $E_t = E_l/4$, shear modulus, $G_{lt} = 0.1 E_l$, Poisson ratios~$\nu_{lt} = 0.2$ and~$\nu_{tl} = \nu_{lt}/4$. The fibre has a width~$w = 0.43 l$ and thickness~$t$. The exact area of the fibre is~$A_0 = 0.43l^2$.

The problem is formulated as a plane stress case and solved with a non-conforming fixed grid by the standard FEM and the LS-XFEM approach. In the standard finite element setting, a finite element is considered to be part of the fibre if the centroid of the finite element is located inside the fibre. Accordingly, these fibres contribute to the finite element stiffness in the FEM solutions, and a slow convergence rate with respect to the chosen element size of the underlying discretization is thus expected for the non-conforming FEM. In the LS-XFEM formalism, a finite element partially covering a fibre still represents that fibre, but the area of the fibre in that finite element is determined using the area integration method  presented in Section~\ref{eq:numerical_integration} and it contributes to the element stiffness accordingly. The results obtained by these two approaches for different grid spacings (element edge lengths) $h$ are presented in Tab.~\ref{infi11} in terms of the (local) fibre stress component~$\sigma_{xx}$ as computed, and the fibre area considered by the numerical model.
\begin{table}
	\centering
	\caption{Computed stresses and areas obtained by the standard non-conforming FEM and LS-XFEM.}
	\label{infi11}
	\renewcommand*{\arraystretch}{1.15}
	\begin{tabular}{c|rrrr}
		$l/h$ & \multicolumn{1}{c}{$\sigma_{xx}^\text{FEM}/\sigma_0$} & \multicolumn{1}{c}{$\sigma_{xx}^\text{LS-XFEM}/\sigma_0$} & \multicolumn{1}{c}{$A^\text{FEM}/A_0$}
		& \multicolumn{1}{c}{$A^\text{LS-XFEM}/A_0$} \\\hline
		$10$ & $1.0750$ & $1.0043$ & $0.9302$ & $0.9957$ \\ 
		$20$ & $1.0750$ & $0.9998$ & $0.9302$ & $1.0000$ \\ 
		$40$ & $1.0117$ & $0.9998$ & $0.9883$ & $1.0000$ \\
	\end{tabular}
\end{table}

As expected, the non-conforming FEM does not yield a good estimate of the stresses with a coarse mesh of 10 elements along the cell edge. On the contrary, the LS-XFEM formalism yields an accurate stress distribution with the same mesh. As the mesh is refined, the non-conforming FEM still struggles to yield the expected accuracy, showing a slow convergence rate.  However, the LS-XFEM formalism captures the geometry with a better accuracy, which also results in a more accurate prediction of the stress level.
%
%----------------------------------
%	BONDED ELEMENTARY NETWORK UNDER TENSION
%----------------------------------
%
\subsubsection{Bonded elementary network under tension}
\begin{figure}
	\centering
	\subfloat[infinitely long fibre network and loading considered]{\includegraphics[height=60mm]{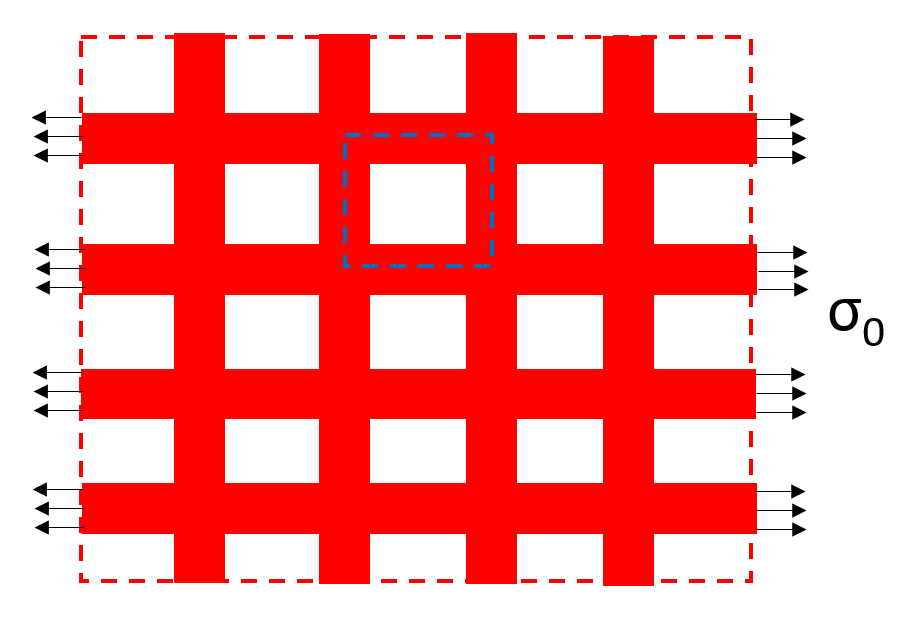}\label{figg4}}
	\hspace{0.5em}
	\subfloat[discretized unit cell with~$ 50\times 50 \times 2 = 5000$ elements non-conforming to the geometry]{\includegraphics[height=60mm]{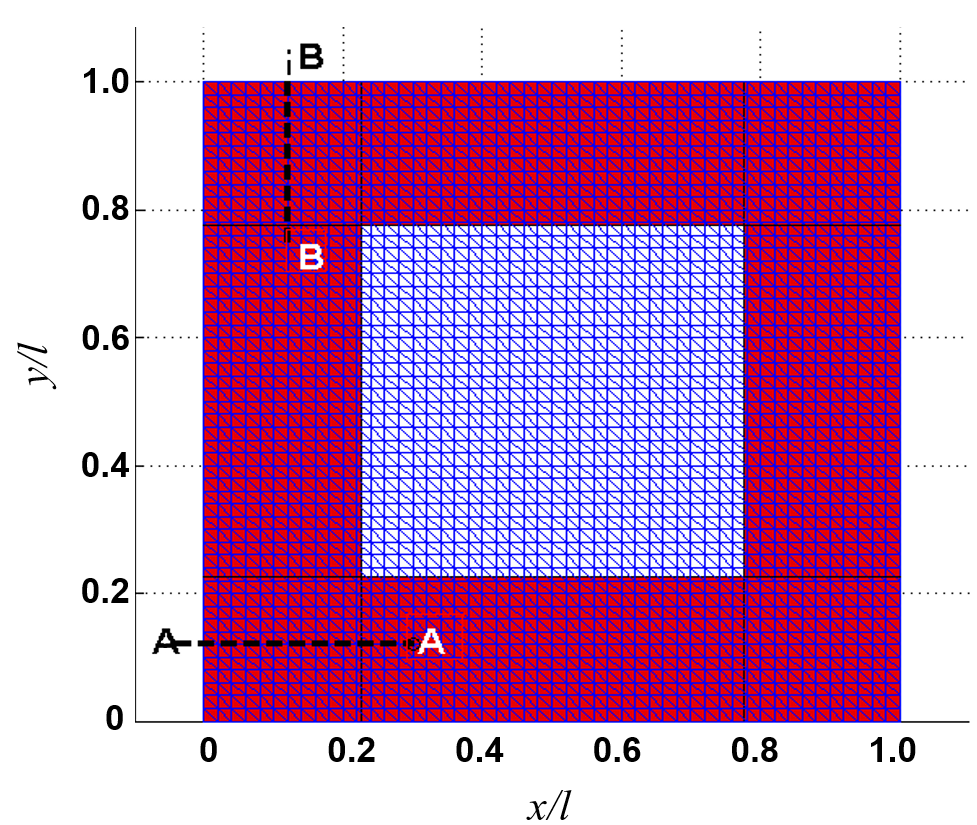}\label{figg3}}
	\caption{(a)~An infinite number of two orthogonal families of parallel, equispaced, infinitely long fibres subjected to a uniaxial stress~$\sigma_0$. (b)~Corresponding periodic unit cell with a regular mesh non-conforming to the geometry. The~$\sigma_{xx}$ profiles along the cross-sections~A--A and~B--B are depicted in Fig.~\ref{cs}.}
	\label{pics:infinite_mesh}
\end{figure}

The attention is next shifted towards the simplest fibre arrangement that involves bonds between two orthogonal families of parallel fibres subjected to a stress~$\sigma_0$, as introduced in~\citep{Bosco1}, see Fig.~\ref{figg4}. Because of periodicity, the unit cell depicted in Fig.~\ref{figg3} is used, with half fibres along the edges. The material parameters are identical to those of the previous example. The LS-XFEM formalism, with a regular mesh with~$h = l/50$ as shown in Fig.~\ref{figg3}, is used to predict the mechanical response, especially focusing on the stress distribution in the bonded regions. This network is also modelled with a very fine FE mesh of size~$400 \times 400 \times 2$ elements conforming to the geometry, serving as a reference solution. It is emphasized that such a conforming mesh is of course only achievable for simple configurations as considered in this example, and is used here mainly to quantify the accuracy of the LS-XFEM method.
\begin{figure}
	\centering
	\subfloat[reference conforming solution~($h = l/400$)]{\includegraphics[trim={2mm 3mm 0mm 0mm},height=60mm]{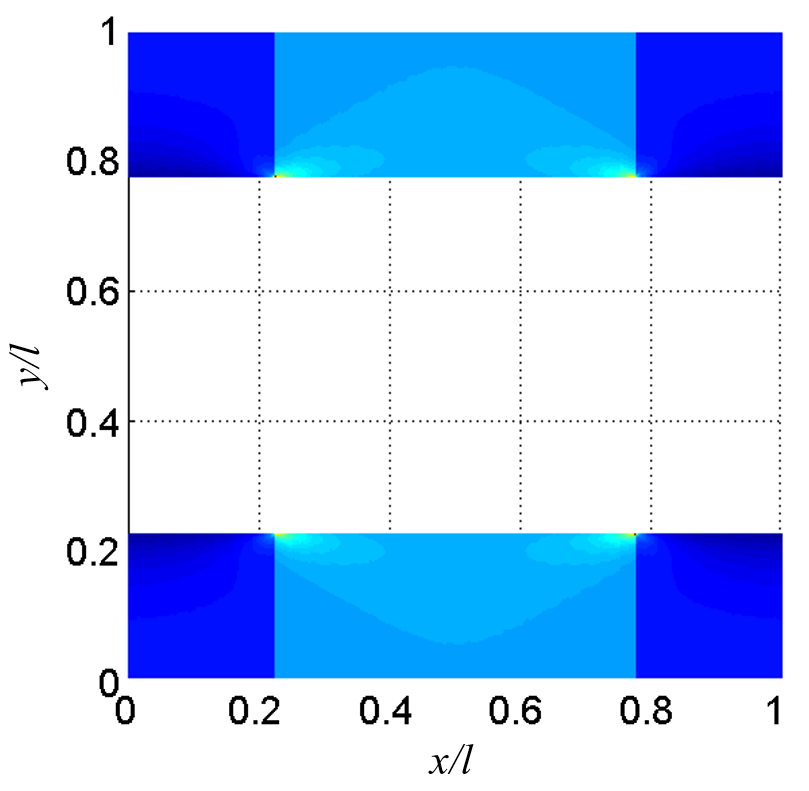}\label{fem400}}
	\hspace{0.5em}
	\subfloat[LS-XFEM ($h = l/50$)]{\includegraphics[trim={2mm 0mm 0mm 0mm},height=60mm]{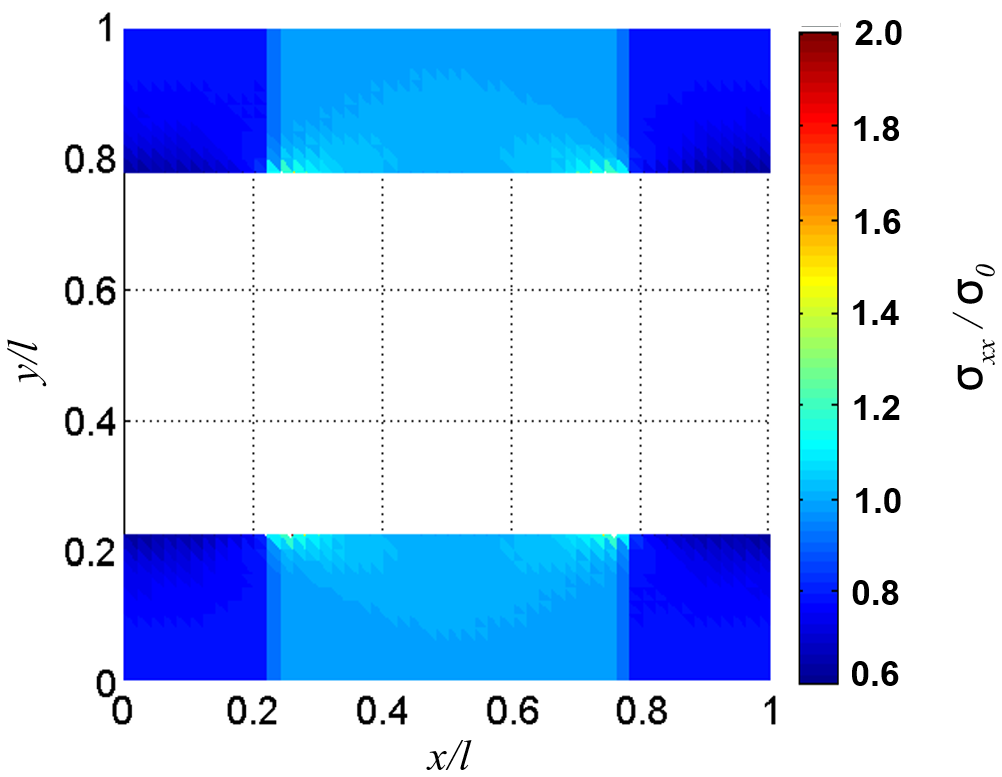}\label{50}}
	\caption{Stress distributions, $\sigma_{xx}/\sigma_0$, in the horizontal fibre.}
	\label{fib1}
\end{figure}
\begin{figure}
	\centering
	\subfloat[reference conforming solution ($h = l/400$)]{\includegraphics[trim={0mm 2mm 0mm 0mm},height=60mm]{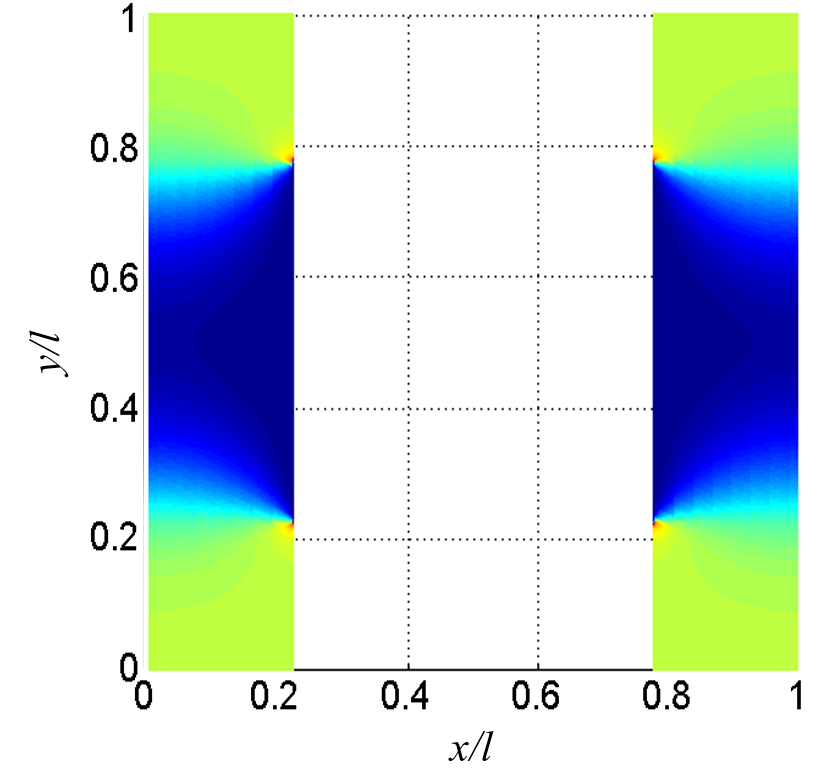}}
	\hspace{0.5em}
	\subfloat[LS-XFEM ($h = l/50$)]{\includegraphics[trim={2mm 1mm 0mm 0mm},height=60mm]{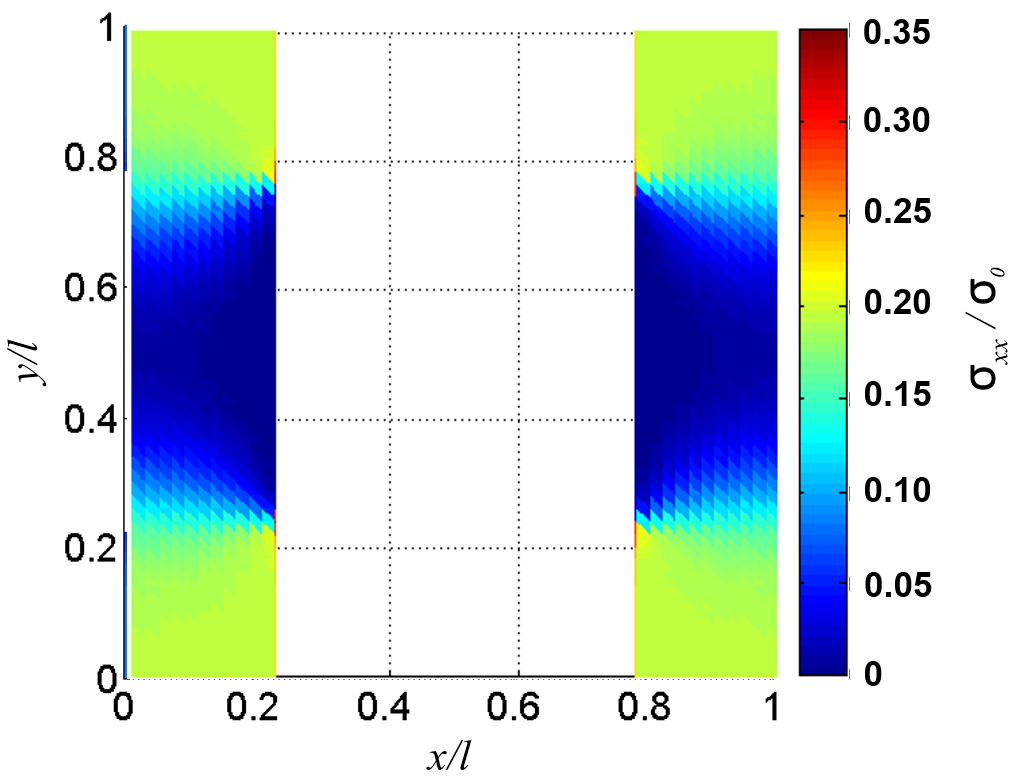}}
	\caption{Stress distributions, $\sigma_{xx}/\sigma_0$, in the vertical fibre.}
	\label{fib2} 
\end{figure}
\begin{figure}
	\centering
	\subfloat[horizontal fibre, cross-section A--A ]{\includegraphics[trim={0mm 0mm 0mm 0mm},height=60mm]{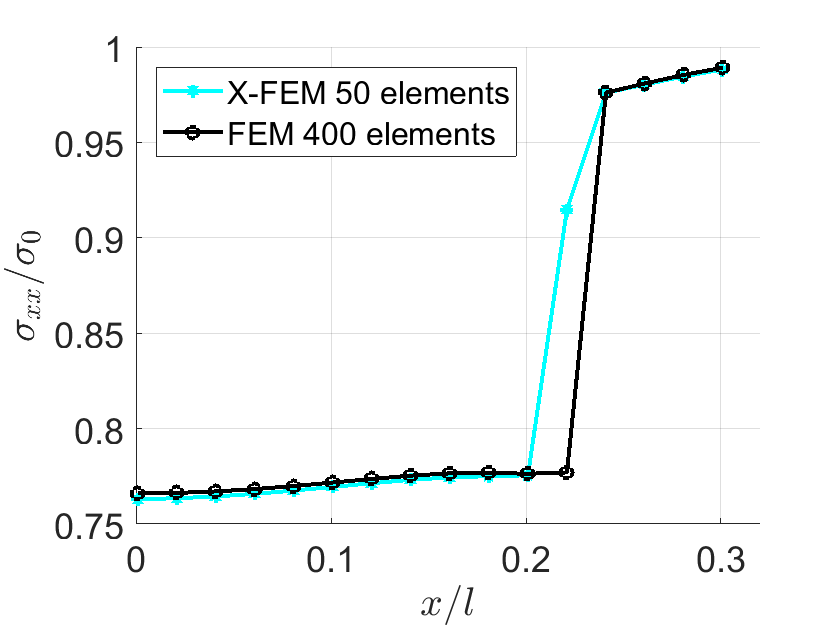}\label{csa}}
	\hspace{0.5em}
	\subfloat[vertical fibre, cross-section B--B]{\includegraphics[trim={0mm 0mm 0mm 0mm},height=60mm]{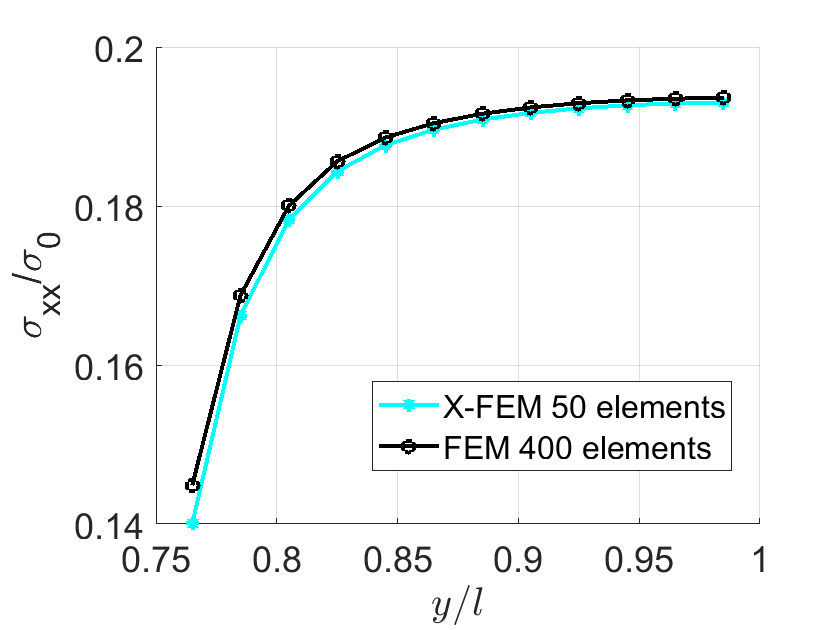}\label{csb}}
	\caption{Stress distributions, $\sigma_{xx}/\sigma_0$, in cross-sections A--A and B--B as defined in Fig.~\ref{figg3} for conforming FEM and LS-XFEM.}
	\label{cs}
\end{figure}

Figs.~\ref{fib1} and~\ref{fib2} show the normalized stress distribution~$\sigma_{xx}$ in the bonded regions for the horizontal and vertical fibres as computed by the LS-XFEM with a non-conforming mesh and the conforming FEM reference solution. The coarse triangular mesh used for the LS-XFEM formalism is evident in the jagged stress distribution in the free standing and bonded regions in Fig.~\ref{50}. Also, these are noticed near the bonded region of the vertical fibres in Fig.~\ref{fib2}. Note that only at locations ($ x/l = 0.2 $), close to the edges of the bond, the stress distribution predicted by LS-XFEM differs from the reference solution with a deviation of 20$\%$, as observed in Fig.~\ref{csa}. This is because with the LS-XFEM formalism employing a geometry non-conforming coarse mesh, the finite element at this location lies in the bond as well as in the free standing fibre. Now, the stress distribution~$\sigma_{xx}$ is high in the free standing fibre resulting in the stress jumps obtained by LS-XFEM. However, in the geometry-conforming mesh employed by the FEM solution, the finite element is lying inside a bond with low~$\sigma_{xx}$. Later on, the stress distribution by the FEM also attains the same value as predicted by LS-XFEM formalism. Most importantly, as noticed in these plots, the normalized stress distributions predicted by the LS-XFEM formalism are qualitatively and quantitatively similar to the reference solutions at other regions, which is further ascertained in the cross-sectional plots through the bonds shown in Fig.~\ref{cs}.

In a fibrous network, the bonded regions are vital for an accurate prediction of the overall response, as well as for the proper reproduction of the local behaviour of the fibres~\citep{Bosco1}. Therefore, the ability of the LS-XFEM formalism to make adequate predictions of the mechanical stress state inside the bond at a lower computational cost than a fine conforming FEM discretization makes it a suitable tool for modelling fibrous networks.
%
%----------------------------------
%	MEDIUM-COMPLEXITY NETWORK
%----------------------------------
%
\subsection{Medium-complexity network}
\label{sec:complex_network}
The attention is now focused on somewhat more complex networks, to illustrate the ability of the proposed formalism to recover information at both the microstructural and macroscopic level. The error committed by the LS-XFEM is quantified by comparison with a reference, finely discretized conforming FEM. The complexity of the networks considered in this section is limited by our ability to generate (and simulate) such a conforming discretization---see Section~\ref{sec:realistic_network} for a complex realistic network, for which this is no longer feasible (but where our approach still works).
%
%----------------------------------
%	NETWORK PARAMETERS
%----------------------------------
%
\subsubsection{Network parameters}
Different network configurations are considered, with coverages of~$c = 0.9$ and~$c = 1.8$. Recall that coverage is defined as the ratio of total area occupied by the fibres in the network to the area of the microstructural unit cell. The characteristic size of the finite elements used in the mesh to model the unit cell is chosen as~$ h_L = l/100 $. After application of the refinement strategy at the fibre edges, the smallest finite element size reduces to $ h_S = l/400 $. For the anisotropic behaviour of the fibres, the material parameters used are identical to those of the previous examples. The coefficients of hygroscopic expansion are taken according to~$ \beta_t = 20\beta_l $ for all cases. A unit change in moisture content, $ \Delta\chi = 1 $, is adopted which is assumed uniform over the entire unit cell.
%
%----------------------------------
%	AVERAGE EXPANSIVITY AND DEFORMED GEOMETRY
%----------------------------------
%
\subsubsection{Average expansivity and deformed geometry}
\begin{figure}
	\centering
	\subfloat[initial network of coverage~$c = 0.9$ and~$q = 0.5$]{\includegraphics[height=0.3\textwidth]{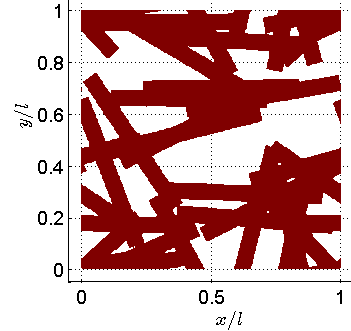}\label{ini}}
	\hspace{0.0em}
	\subfloat[refined mesh for LS-XFEM]{\includegraphics[height=0.3\textwidth]{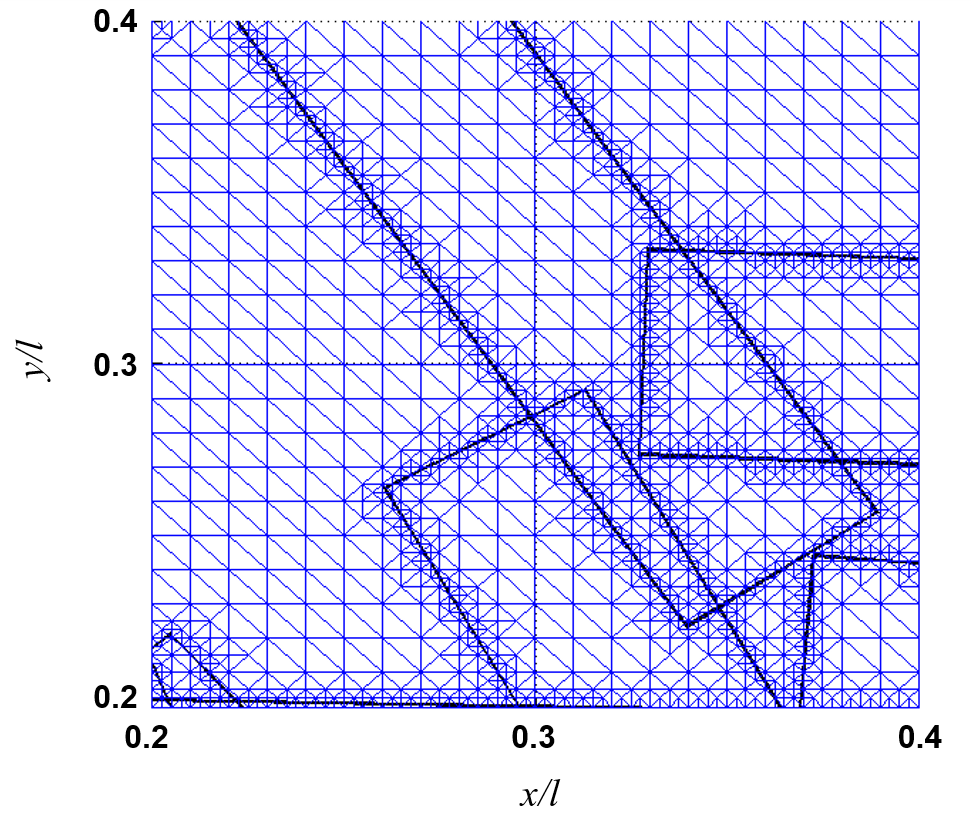}\label{refmesh}}
	\hspace{0.0em}	
	\subfloat[conforming mesh for FEM]{\includegraphics[height=0.3\textwidth]{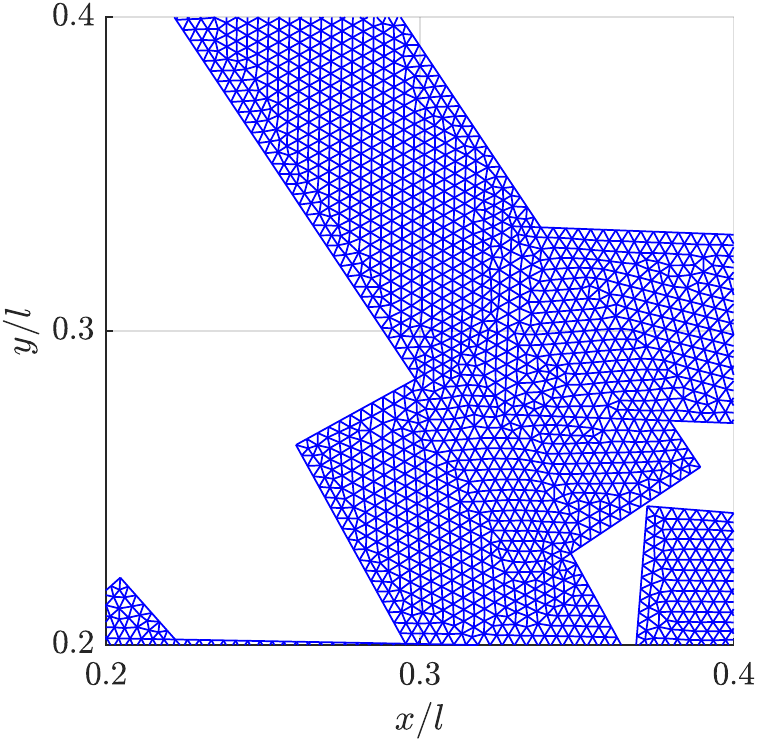}\label{FEMc}}
	\caption{Geometry of the complex network and the magnified view of meshes, coverage~$c = 0.9$ and anisotropy parameter~$q = 0.5$.}
	\label{fig:complex_geometry}
\end{figure}

\begin{figure}
	\centering
	\subfloat[reference conforming FEM with~$h_C$ ($50373$ nodes)]{\includegraphics[height=59mm]{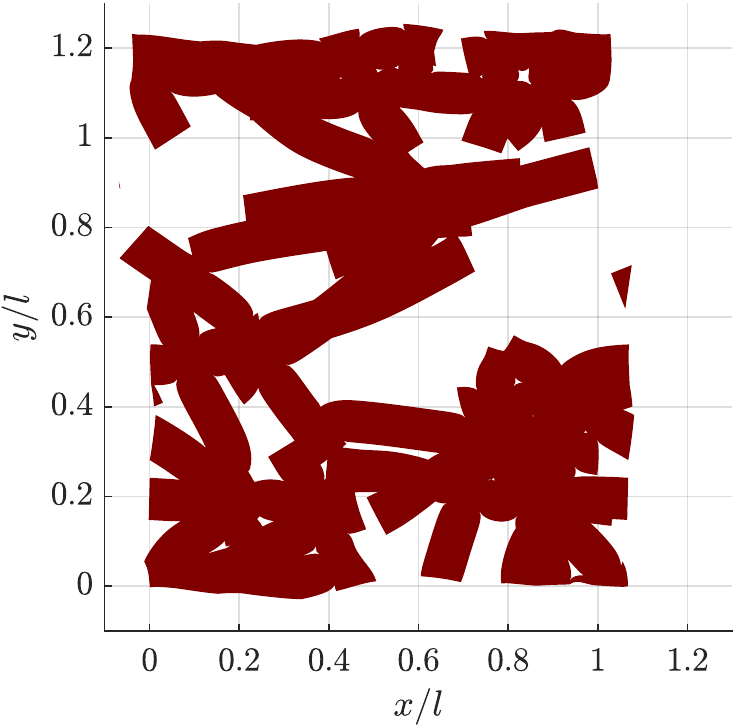}\label{FEMdef}}
	\hspace{0.5em}
	\subfloat[LS-XFEM with refined mesh ($48475$ nodes)]{\includegraphics[height=60mm]{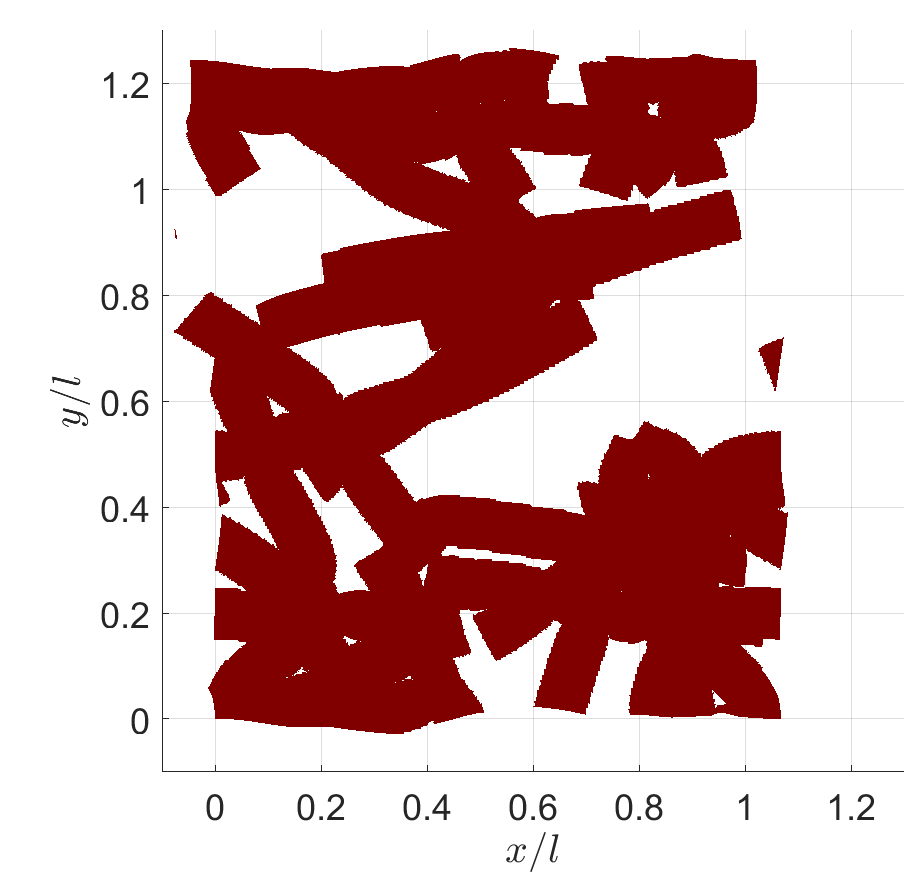}\label{XFEMd}}
	\caption{Deformed network as computed by the conforming FEM and LS-XFEM. The displacements have been magnified by a factor of~$50$.}
	\label{fig:deformed}
\end{figure}

The initial anisotropic network for coverage~$c=0.9$ and anisotropy parameter~$q = 0.5$ is shown in Fig.~\ref{ini}. The hygro-mechanical response of the unit cell is computed by solving the static equilibrium problem for a unit change in moisture content, $ \Delta\chi $. This generates a hygroscopic load causing deformation in the network, which is computed by means of the LS-XFEM formalism on a discretization with four levels of refinements, see Fig.~\ref{refmesh}. The response of the same network is computed as well using a conforming FEM with the maximum element size~$h_C = l/250$, the discretization of which is shown in Fig.~\ref{FEMc}. For the conforming discretization, the Gmsh mesh generator, version~4.5.6, has been used~\citep{Geuzaine2009}. It can be observed in Fig.~\ref{fig:deformed} that the deformed geometry obtained by the LS-XFEM (with a relatively coarse mesh), in Fig.~\ref{XFEMd}, is similar to the deformed network obtained by the conforming FEM (with a very fine mesh), in Fig.~\ref{FEMdef}.

\begin{table}
	\centering
	\caption{Computed effective hygro-expansive coefficients normalized with~$\beta_l$ for the anisotropic network ($c = 0.9$ and~$q = 0.5$).}
	\label{tab1}
	\renewcommand*{\arraystretch}{1.15}
	\begin{tabular}{r|rrr}
		\multicolumn{1}{c|}{Approach} & \multicolumn{1}{c}{$\overline{\beta}_{xx}/\beta_l$} & \multicolumn{1}{c}{$\overline{\beta}_{yy}/\beta_l$} & \multicolumn{1}{c}{$\overline{\beta}_{xy}/\beta_l$} \\\hline
		conf. FEM & $2.3927$ & $8.2062$ & $-1.3936$ \\
		LS-XFEM & $2.4278$ & $8.6910$ & $-1.6682$		
	\end{tabular}
\end{table}

\begin{figure}
	\centering
	\includegraphics[clip,height=60mm]{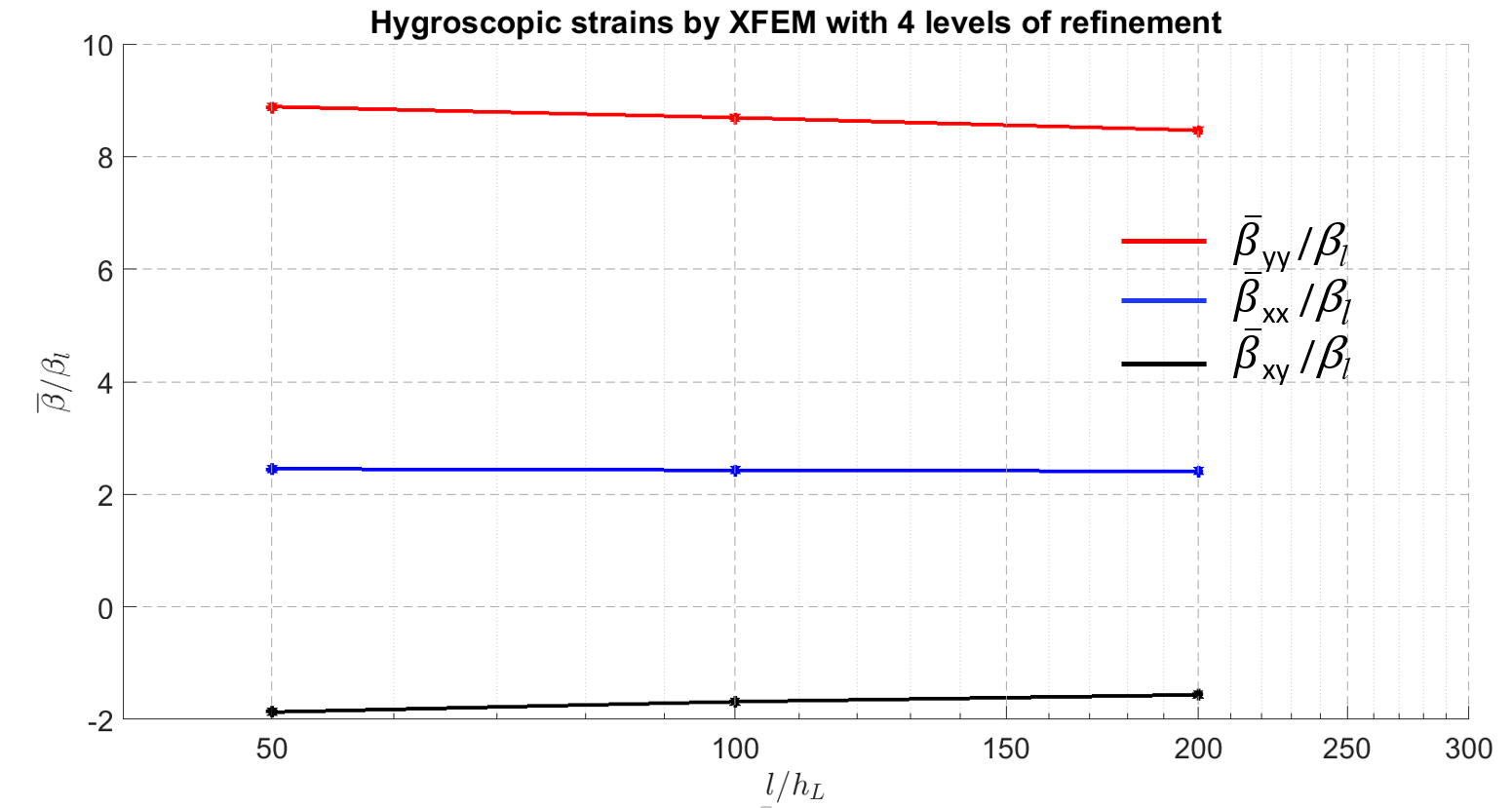}
	\caption{Convergence of the effective hygroscopic coefficients normalized with~$\beta_l$ obtained by the LS-XFEM as a function of the characteristic element size.} 
	\label{convi11}
\end{figure}

This is further illustrated by the overall behaviour on the basis of the computed overall hygroscopic coefficients of the network, as listed in Tab.~\ref{tab1}. The anisotropic network fabric ($q = 0.5$) causes a pronounced overall anisotropy: the expansivity $\overline{\beta}_{yy}$ in the cross direction exceeds that of the machine direction, $\overline{\beta}_{xx}$, by more than a factor of three. This is due to the fact that the fibres are on average more oriented in the machine direction, i.e. their expansion occurs predominantly in the cross direction. The values obtained by the LS-XFEM have a relative deviation of less than $1.5\%$ for~$\overline\beta_{xx}$ and~$6\%$ for~$\overline\beta_{yy}$, when compared against the conforming FEM solution for the same network. In Fig.~\ref{convi11}, convergence of the normalized effective hygro-expansivity of the network with an increasing number of elements in the mesh for the LS-XFEM formalism is shown. There is no significant change in the effective coefficients of the network with  an increase in the number of elements. Even with the coarsest discretization considered, i.e.~$h_L = l/50$, we obtain a response which is within~$10\%$ of the response obtained at~$h_L = l/200$.

\begin{figure}
	\centering
	\subfloat[initial network]{\includegraphics[clip,height=65mm]{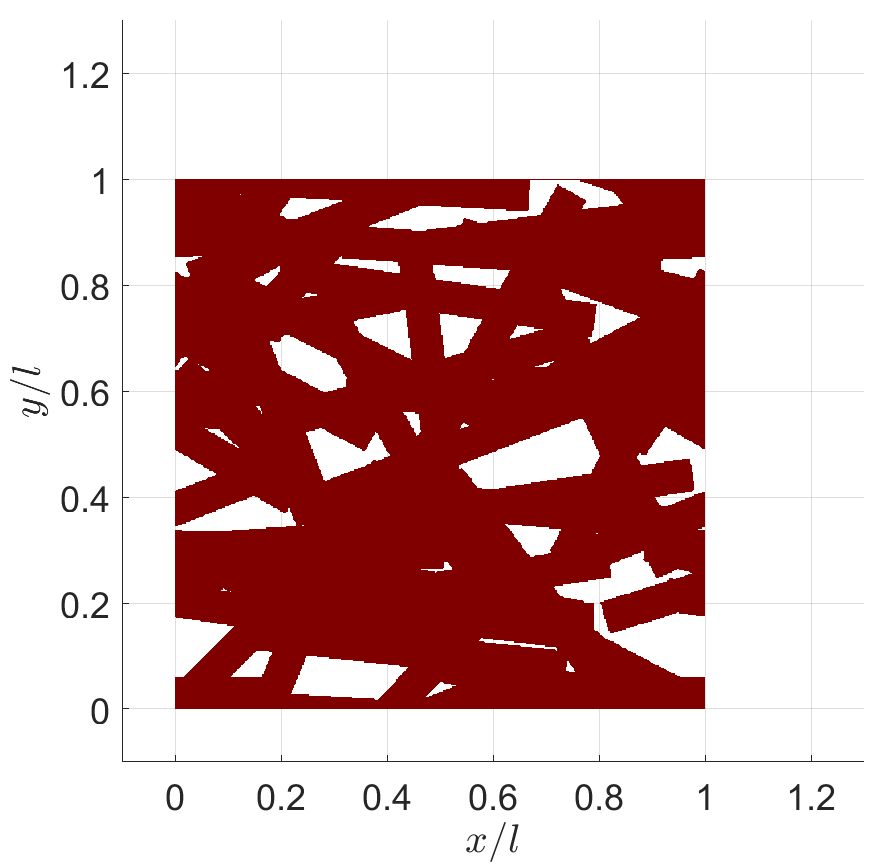}\label{highc}}
	\hspace{0.5em}
	\subfloat[deformed network, LS-XFEM]{\includegraphics[clip,height=65mm]{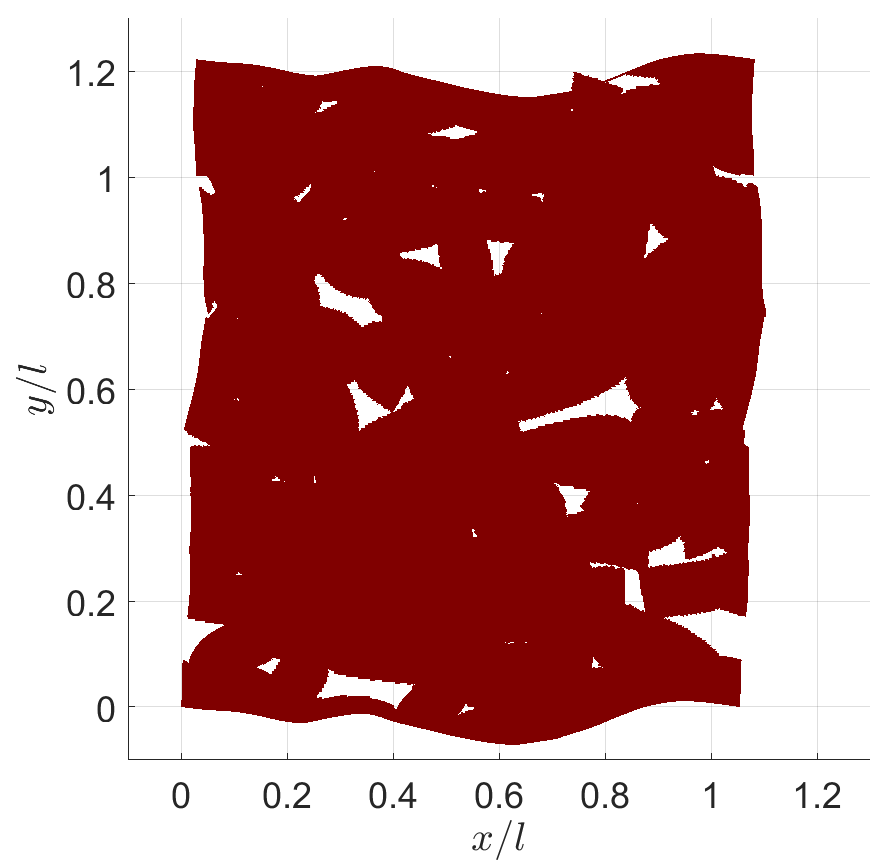}}
	\caption{A high coverage network ($c = 1.8$ and~$q = 0.5$). The displacements have been magnified by a factor of~$50$.}
	\label{highc1}
\end{figure}

In Fig.~\ref{highc1}, an anisotropic network with a higher coverage is considered ($c = 1.8$). For this case, the bonded area in the network is larger, resulting in a comparatively higher hygroscopic strain and overall deformation. As in the previous case, the anisotropic orientation of fibres in the network results in a higher expansion along the cross direction. Hence, both the anisotropy and higher coverage contribute to a higher expansion in the cross direction when compared with networks having low coverages.

Finally, an isotropic network ($q = 0$) with low coverage of~$c = 0.9$ is considered. The values of the expansivity in machine direction ($\overline{\beta}_{xx}/\beta_l$) and cross direction ($\overline{\beta}_{yy}/\beta_l$) are listed in Tab.~\ref{tab2}. Theoretically, if the network would be truly isotropic and representative, the listed values for the machine and cross directions should be identical (i.e.~$\overline{\beta}_{xx} = \overline{\beta}_{yy}$) and the shear component should vanish (i.e.~$\overline{\beta}_{xy} = 0$). However, $\overline{\beta}_{xx}$ and~$\overline{\beta}_{yy}$ differ by approximately~$10\%$ and~$\overline{\beta}_{xy} \neq 0$ due to the sparsity and the statistically small size of the network used. The LS-XFEM agrees within~$6.5\%$ with the reference conforming FEM results, indicating an adequate approximation of the hygroscopic properties of the considered network.
\begin{table}[t]
	\caption{Computed effective hygro-expansive coefficients normalized with~$\beta_l$ for the isotropic network ($c = 0.9$ and~$q = 0$).}
	\label{tab2}
	\renewcommand*{\arraystretch}{1.15}
	\centering
	\begin{tabular}{c|rrr} 
		Approach & \multicolumn{1}{c}{$\overline{\beta}_{xx}/\beta_l$} & \multicolumn{1}{c}{$\overline{\beta}_{yy}/\beta_l$} & \multicolumn{1}{c}{$\overline{\beta}_{xy}/\beta_l$} \\\hline
		conf. FEM & $2.7943$ & $3.0957$ & $1.2783$ \\
		LS-XFEM & $2.7989$ & $3.1457$ & $1.3573$		
	\end{tabular}
\end{table}
%
%----------------------------------
%	LOCAL BEHAVIOUR
%----------------------------------
%
\subsubsection{Local behaviour}
\begin{figure}
	\centering
	\subfloat[normalized strain distribution, $\varepsilon_{xx}/(\beta_l\Delta\chi)$]{\includegraphics[scale=0.95]{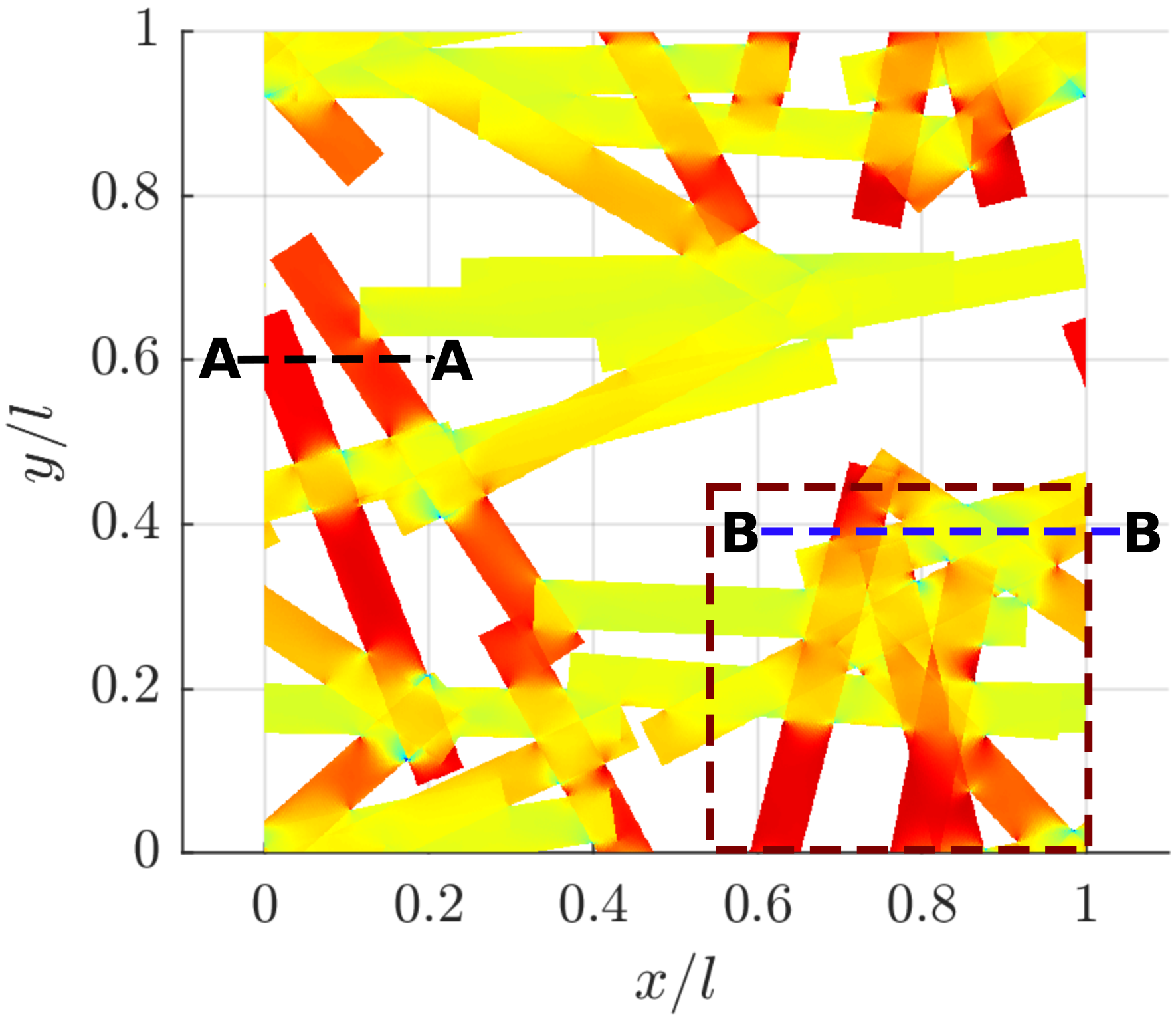}\label{FEMconf}}
	\hspace{2.5em}
	\subfloat[magnified view of the dashed region]{\includegraphics[scale=0.93]{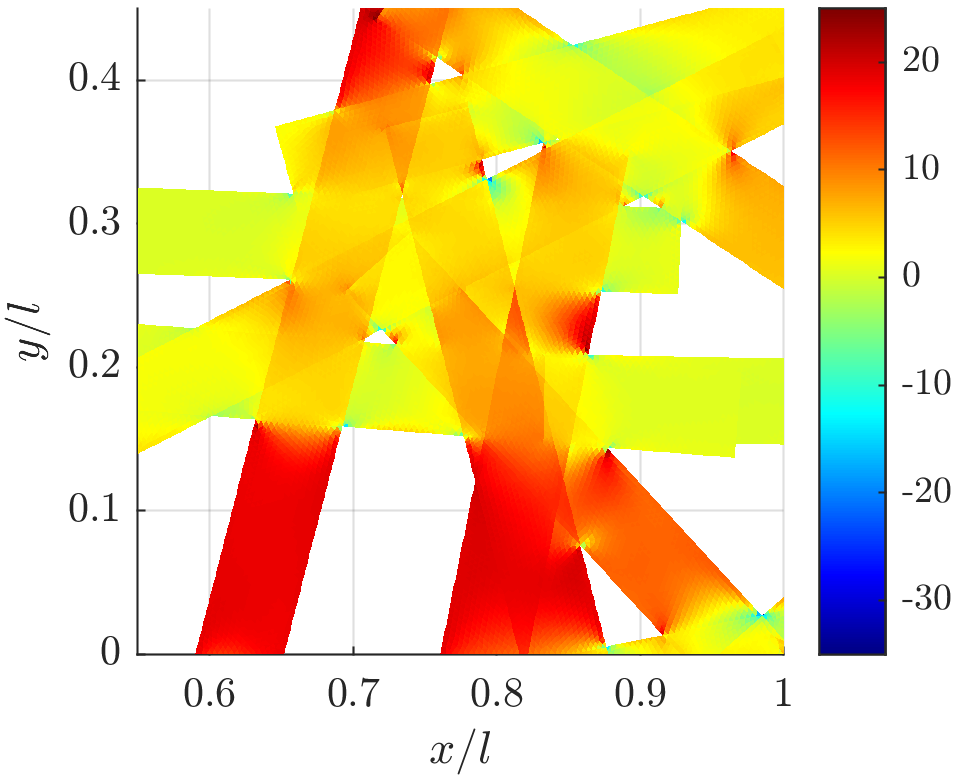}\label{FEMconf1}}
	\caption{Reference conforming FEM solution for medium-complexity network with mesh size~$h_C = l/250$ ($50373$ nodes, $c = 0.9$, and~$q = 0.5$).}
	\label{FEMr3}
\end{figure}

\begin{figure}
	\centering
	\subfloat[normalized strain distribution, $\varepsilon_{xx}/(\beta_l\Delta\chi)$]{\includegraphics[height=60mm]{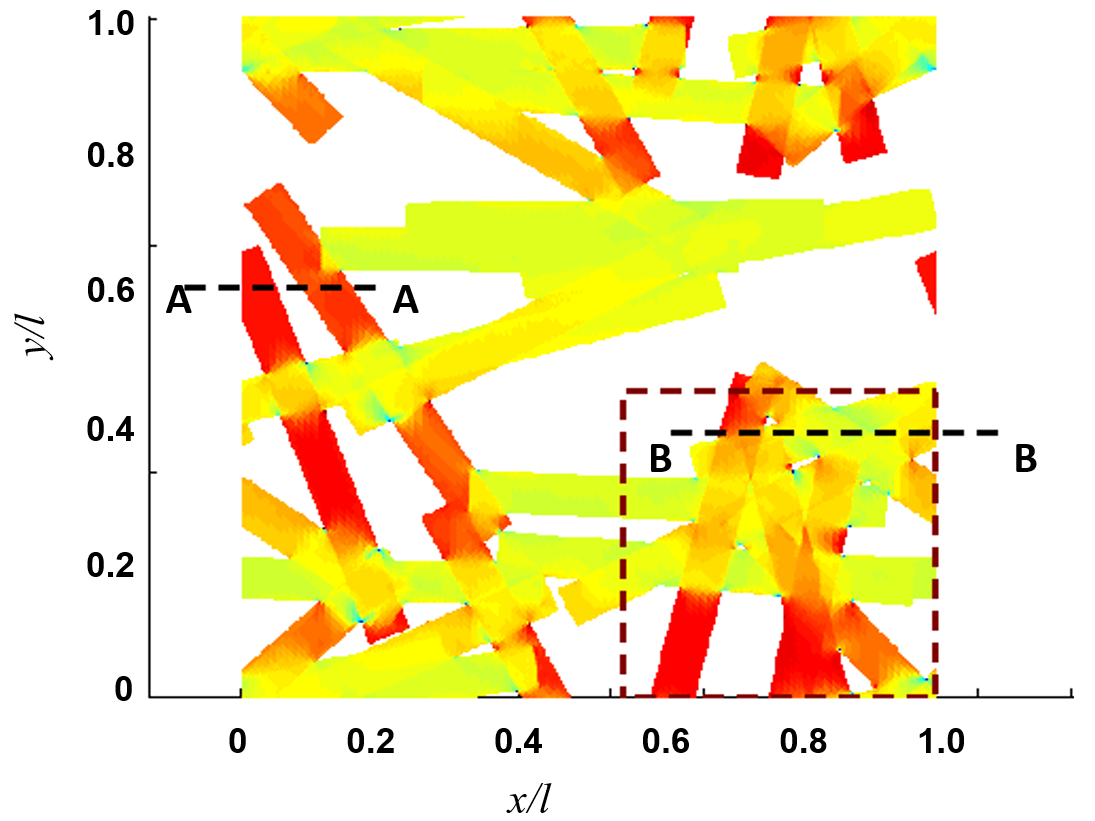} \label{XFEMr}}
	\hspace{0.0em}
	\subfloat[magnified view of the dashed region]{\includegraphics[clip,height=60mm]{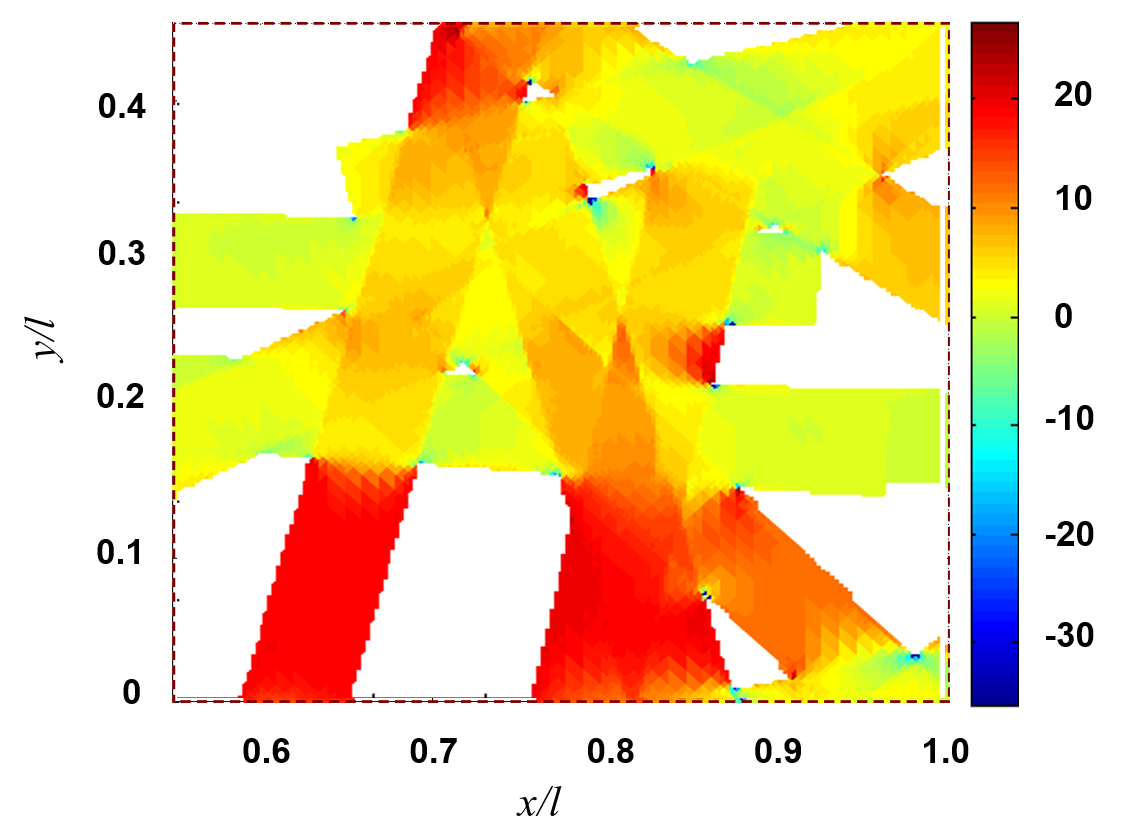}\label{XFEMr1}}
	\caption{LS-XFEM solution for medium-complexity network obtained with refined mesh ($48475$ nodes, $c = 0.9$, and~$q = 0.5$).}
	\label{XFEMr2}
\end{figure}

Local strain distributions in the medium-complexity fibrous networks obtained by the LS-XFEM formalism are analysed in this section by comparing them against the high resolution conforming FEM solutions. In Figs.~\ref{FEMconf} and~\ref{XFEMr}, the normalized strain distributions are plotted for, respectively, a conforming FEM mesh ($50373$ nodes) and the LS-XFEM ($48475$ nodes). The strain distribution obtained by the LS-XFEM is accurately computed in the overall network as well as in the bonds. For a better comparison, the magnified views of the normalized strain distribution are plotted in Figs.~\ref{FEMconf1} and~\ref{XFEMr1}.
\begin{figure}
	\centering
	\includegraphics[scale=1]{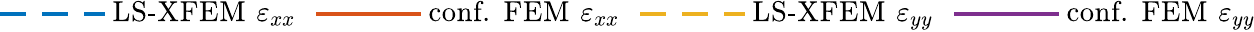}\\
	\subfloat[cross-section A--A]{\includegraphics[height=60mm]{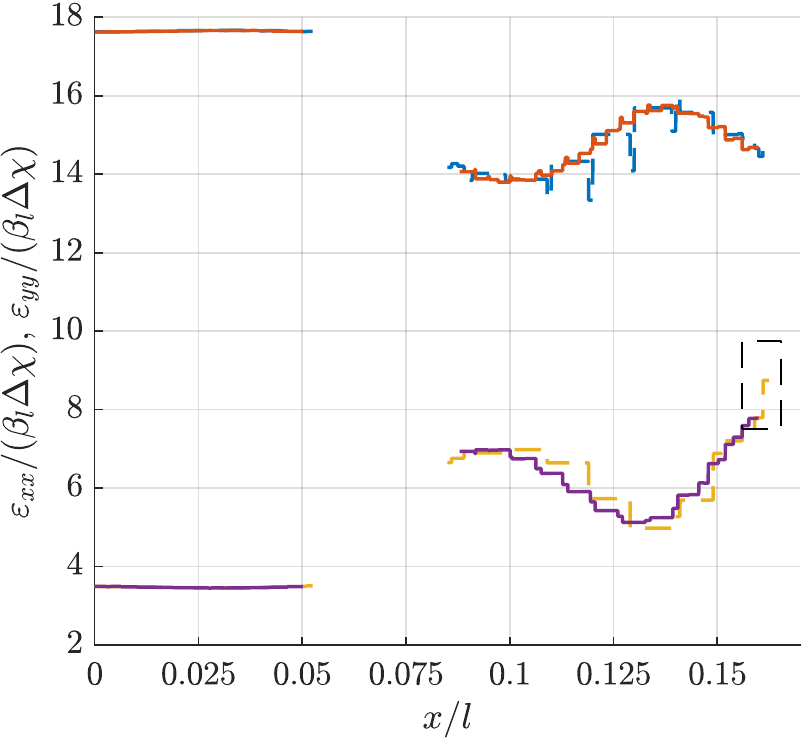}\label{cs2}}
	\hspace{1.0em}
	\subfloat[cross-section B--B]{\includegraphics[height=60mm]{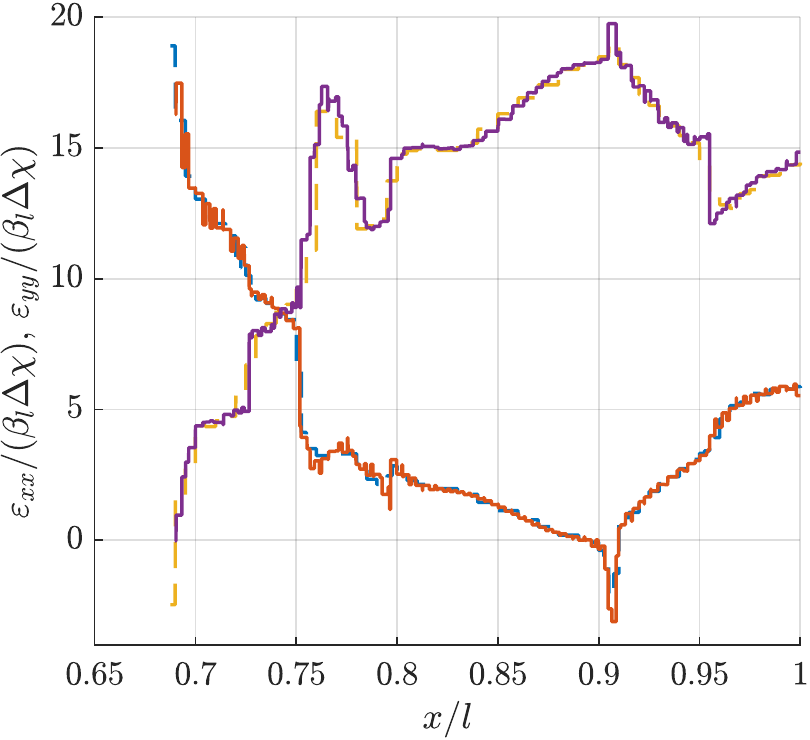}\label{cs3}}
	\caption{Normalized strain in the network at cross-sections A--A and B--B.}
	\label{css2}
\end{figure}
\begin{figure}
	\centering
	\subfloat[conforming FEM ($50373$ nodes)]{\includegraphics[trim={0mm 0mm 0mm 0mm},height=65mm]{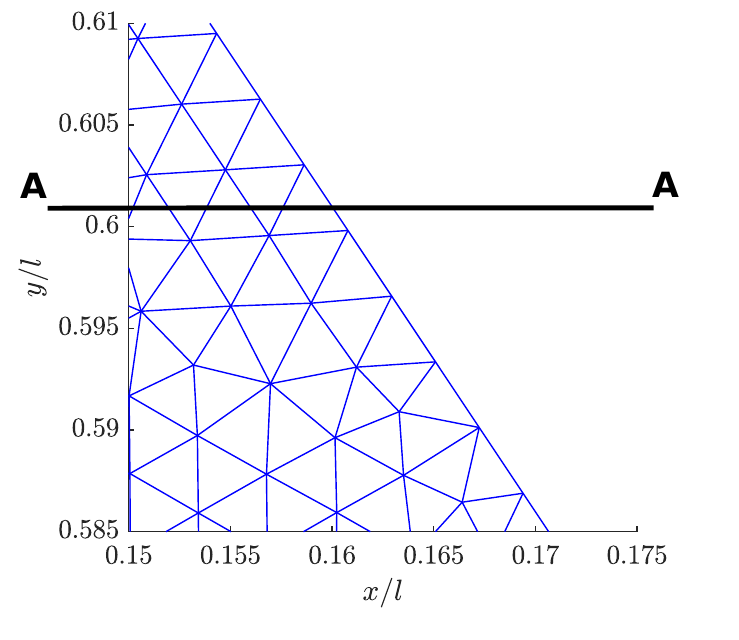}\label{FEMmes}}
	\hspace{0.5em}
	\subfloat[LS-XFEM formalism ($48475$ nodes)]{\includegraphics[trim={0mm 0mm 0mm 0mm},height=65mm]{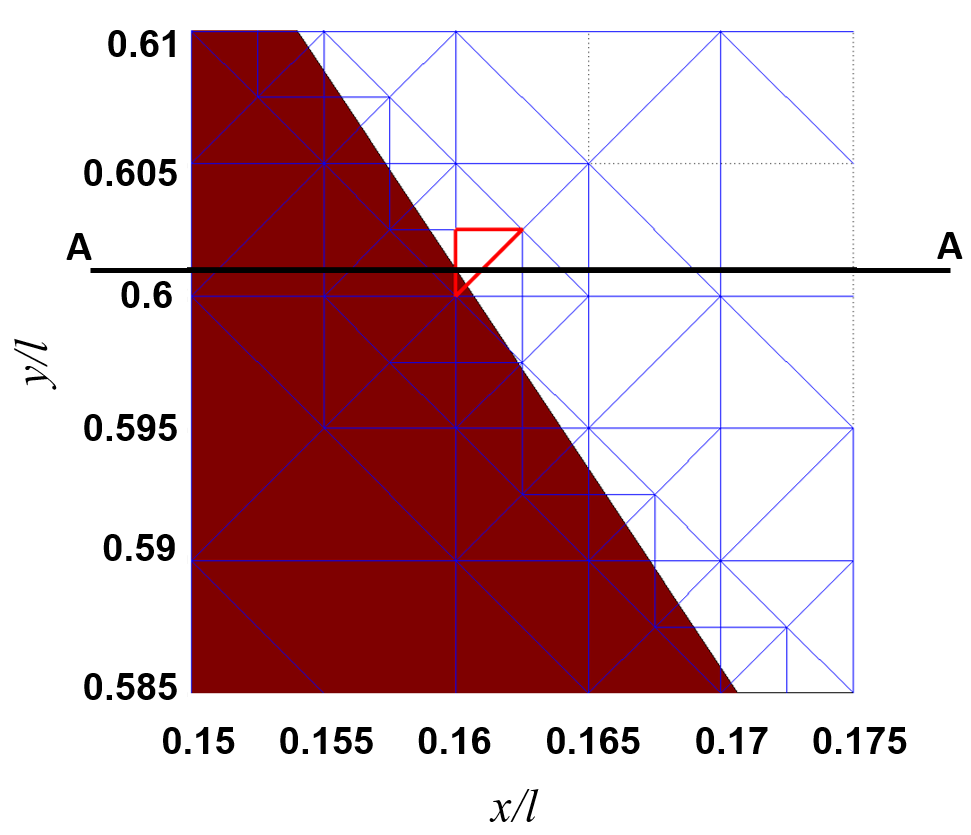}\label{XFEMmes}}
	\caption{A magnified view of the discretized network at cross-section A--A. An interfacial LX-XFEM element located close to~$x=0.16$ is highlighted.}
\end{figure}

The normalized strains along the cross-section A--A and B--B are further plotted in Fig.~\ref{css2}. At a few locations, e.g.~close to $x/l = 0.16$ in the dashed box in Fig.~\ref{cs2}, a strain value is predicted by LS-XFEM, whereas there is no value for the conforming FEM. This is because there is a void for the conforming FEM, whereas in the LS-XFEM solution, the corresponding finite element is identified as an interfacial element, as is noticeable in Fig.~\ref{XFEMmes}. Although there is a strain value predicted inside the entire interfacial element, the area of fibre in that element is captured accurately through the sub-triangulation stated earlier. Similar effects would occur also in bonded areas.
%
%----------------------------------
%	COMPLEX REALISTIC NETWORK
%----------------------------------
%
\subsection{Complex realistic network}
\label{sec:realistic_network}
In the last example, the capabilities of the proposed LS-XFEM methodology are demonstrated on the realistic fibre network of Fig.~\ref{nett}, for which we were unable to obtain a conforming triangulation within a reasonable computing time due to its complex geometry. Several thousands of densely intersecting and overlapping fibres are present within the employed periodic cell, which proved to pose a challenge even for state-of-the-art mesh generators such as Gmsh. The individual fibres are of length~$l_f = 0.5l$ and width~$w_f = l_f/25$, and the material parameters used for the anisotropic behaviour of fibres, their hygroscopic expansion, as well as the integration tolerance of the LS-XFEM method are identical to those of the previous examples. A unit change in moisture content, $ \Delta\chi = 1 $, is adopted, which is again assumed uniform over the entire unit cell. Because of the high density of the network, a uniform mesh with~$200 \times 200$ nodes is used.

\begin{figure}
	\centering
	\subfloat[normalized strain distribution, $\varepsilon_{xx}/(\beta_l\Delta\chi)$]{\includegraphics[height=65mm]{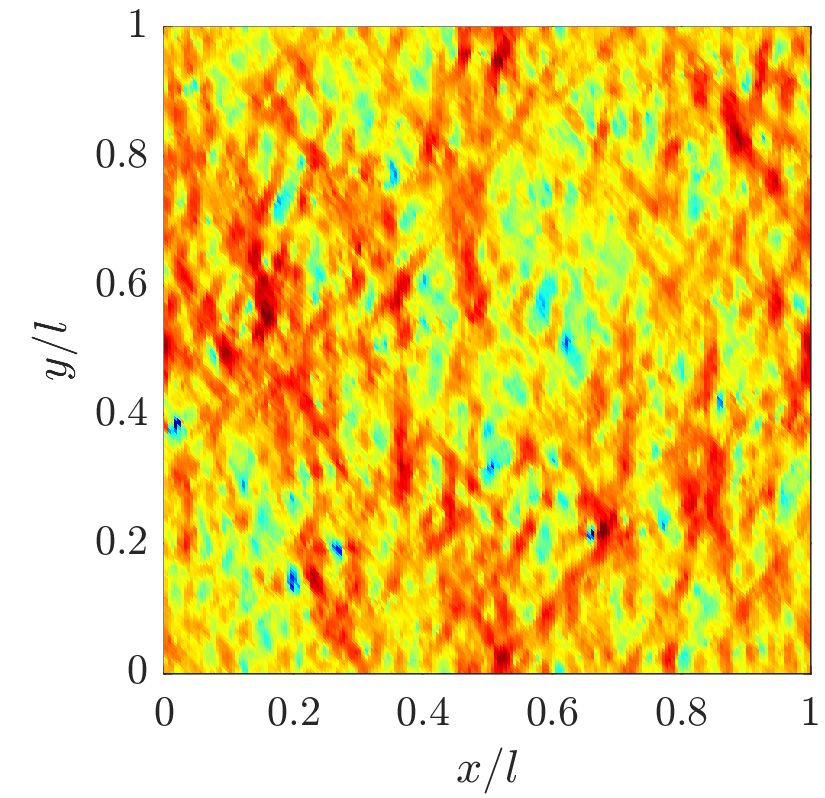}\label{fig:realXFEMa}}
	\hspace{0.5em}
	\subfloat[normalized strain distribution, $\varepsilon_{yy}/(\beta_l\Delta\chi)$]{\includegraphics[clip,height=65mm]{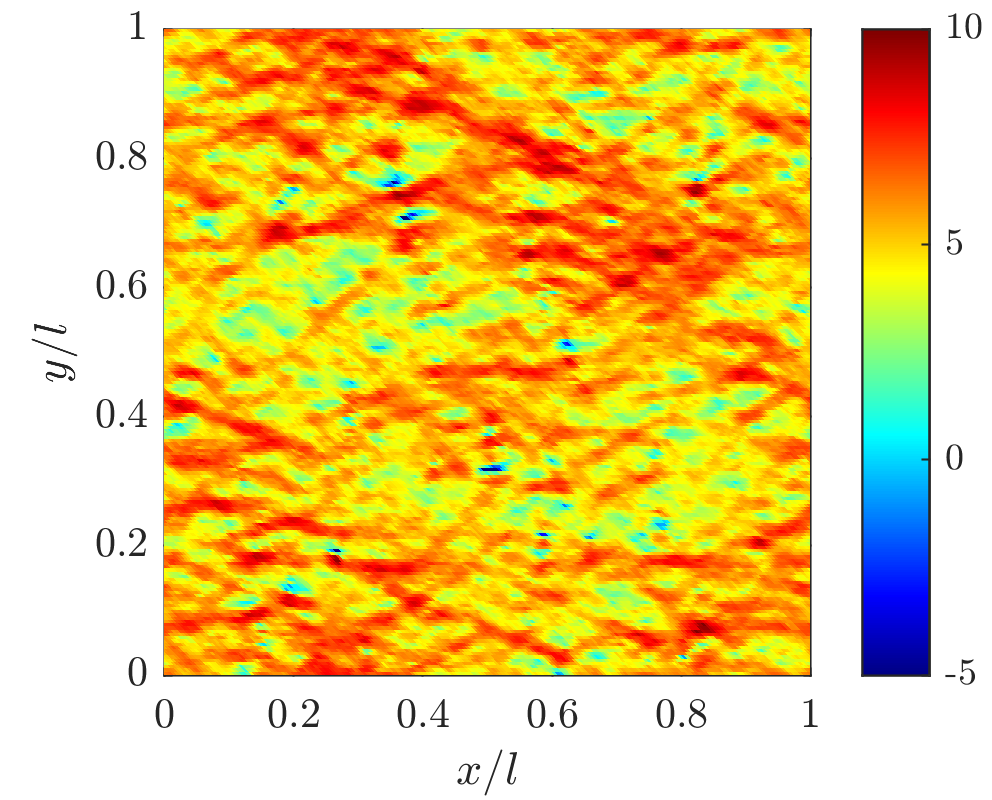}\label{fig:realXFEMb}}
	\caption{LS-XFEM solution for the realistic network shown in Fig.~\ref{nett}, obtained with a regular mesh ($40401$ nodes, $c = 10$, and~$q = 0$).}
	\label{fig:realXFEM}
\end{figure}

The resulting effective hygro-expansive coefficients are summarized in Tab.~\ref{tab:real}. The coefficients significantly more accurately satisfy the conditions imposed by isotropy (i.e.~$\overline{\beta}_{xx} = \overline{\beta}_{yy}$ and~$\overline{\beta}_{xy} = 0$), as compared to Tab.~\ref{tab2}. This is expected because the considered cell contains many more fibres, sampling their orientation distribution much more closely. The remaining discrepancies are due to the relatively small periodic cell size used, of twice the fibre length. Local distributions of strains obtained by the LS-XFEM are shown in Fig.~\ref{fig:realXFEM}, reflecting a much more complicated spatial dependence as compared to the medium-complexity network of Fig.~\ref{XFEMr}.

The presented results demonstrate the ability of the proposed LS-XFEM formalism to resolve the effect of the complexity of such a dense network on the hygro-mechanical behaviour at the sheet-scale. With the LS-XFEM approach, it is thus possible to systematically study the effect of different microstructural parameters of realistic networks with high coverages at an affordable computational cost.
\begin{table}
	\caption{Computed effective hygro-expansive coefficients normalized with~$\beta_l$ for the realistic isotropic network shown in Fig.~\ref{nett} ($c = 10$ and~$q = 0$).}
	\label{tab:real}
	\renewcommand*{\arraystretch}{1.15}
	\centering
	\begin{tabular}{c|rrr} 
		Approach & \multicolumn{1}{c}{$\overline{\beta}_{xx}/\beta_l$} & \multicolumn{1}{c}{$\overline{\beta}_{yy}/\beta_l$} & \multicolumn{1}{c}{$\overline{\beta}_{xy}/\beta_l$} \\\hline
		LS-XFEM & $5.0424$ & $5.2602$ & $0.0329$
	\end{tabular}
\end{table}
%
%-----------------------------------------------------------------------------
%	CONCLUSIONS
%-----------------------------------------------------------------------------
%
\section{Conclusions}
\label{sec:conclusions}
This paper presented a computational methodology for analysis of the hygro-mechanical response of fibre networks using an XFEM framework with a level set based geometry description. A two-dimensional unit cell consisting of a network of fibres has been considered to this end. It was subjected to free expansion by a uniform field of hygroscopic loading caused by a change in moisture. Rectangular fibres were generated randomly to produce periodic cells representing complex networks. The initial geometry of each fibre was described using a level-set function. The coupling between level-sets and the XFEM was carried out by means of a Heaviside function enrichment based on the level-set function. The edges of the fibres were captured from the nodal values of the level-set function. In this manner, a connection was established between the finite element mesh and the internal fibre geometry using the level-set formalism, which simplifies and improves the efficiency of computations for complex geometries in XFEM.

An accurate tracking of the boundaries of the fibres in the network was achieved, especially in the bonded regions, which matter for the prediction of the hygro-mechanical response of the network subjected to moisture infiltration. Local fields in the network (stress, strains and displacements) could be evaluated accurately as compared to a reference finite element solution with a fine conforming mesh. The LS-XFEM results were obtained with a relatively small system size, whereas a comparably accurate non-conforming FEM simulation required a significantly higher number of elements and degrees-of-freedom. It has been demonstrated, on a realistic network, that the LS-XFEM approach is capable of predicting hygro-expansive behaviour of high-coverage networks, for which we were unable to obtain a conforming triangulation within a reasonable amount of computing time. The LS-XFEM formalism is thus a suitable tool for modelling of realistic and high-coverage fibrous networks while capturing the interfaces with an adequate accuracy.

Possible extensions of this work include the modelling of irreversible shrinkage behaviour in paper-like materials. During the manufacturing process of paper, internal stresses are developed when the paper is dried under tension. They are released when the paper is subjected to a moisture cycle, resulting in irreversible shrinkage~\citep{Bosco3}. This behaviour can be included by considering suitable constitutive models.
%
%-----------------------------------------------------------------------------
%	APPENDIX A
%-----------------------------------------------------------------------------
% \appendix
%
% \section{Appendix A}
% \label{Sect:A}
%
%-----------------------------------------------------------------------------
%	ACKNOWLEDGEMENTS
%-----------------------------------------------------------------------------
%
\section*{Acknowledgements}
The first author would like to acknowledge the financial support granted by the European Commision, grant agreement \textnumero~2013-0043, as well as Materials Innovation Institute (M2i) and Canon Production Printing.
%
%-----------------------------------------------------------------------------
%	REFERENCES
%-----------------------------------------------------------------------------
%
%\section*{References}
%
\bibliography{mybibfile}
\end{document}